\documentclass[a4paper]{scrartcl}

\usepackage[dvips]{graphicx, color}
\usepackage{bbm}
\usepackage[english]{babel}
\usepackage[babel]{csquotes} 
\usepackage[super]{natbib}
\usepackage{amssymb,amsmath}
\usepackage[auth-sc]{authblk}
\usepackage{enumerate}
\usepackage{epstopdf}
\usepackage{float}
\def\prodint{\mathop{\ensuremath{\lower0.4ex\hbox{\huge $\mathit{\pi}$}}}}
\usepackage{booktabs}
\usepackage{tabularx}
\usepackage[table]{xcolor}
\usepackage{rotating}
\usepackage{multirow}
\usepackage[margin=10pt,font=footnotesize,labelfont=bf]{caption}

\usepackage{tikz}
\usetikzlibrary{calc}
\usetikzlibrary{positioning}
\usetikzlibrary{arrows}

\usepackage{hyperref}

\usepackage{threeparttable}

\DeclareNewTOC[type=legend,nonfloat]{lgd}

\newcommand*{\legend}[1]{
  \begingroup
    \setkomafont{caption}{\footnotesize}
    \let\captionformat\relax
    \captionbelowof{legend}{#1}
  \endgroup
}

\def\Bl{\Bigl}
\def\Br{\Bigr}

\newcommand\pfs{\mathrm{PFS}}
\newcommand\os{\mathrm{OS}}
\newcommand{\sos}{S_{\mathrm{OS}}}
\newcommand{\spfs}{S_{\mathrm{PFS}}}
\newcommand{\hos}{h_\mathrm{OS}}
\newcommand{\hpfs}{h_\mathrm{PFS}}
\newcommand{\hros}{\mathrm{HR}_{\mathrm{OS}}}
\newcommand{\hrpfs}{\mathrm{HR}_{\mathrm{PFS}}}
\newcommand{\dif}{\mathrm{\;d}}
	 
\begin{document}

\title{Oncology clinical trial design planning based on a multistate model that jointly models progression-free and overall survival endpoints}
\author[1]{Alexandra Erdmann}
\author[1]{Jan Beyersmann}
\author[2]{Kaspar Rufibach}

\affil[1]{Institute of Statistics, Ulm University, Ulm, Germany}
 
 \affil[2]{Product Development Data Sciences,
F. Hoffmann-La Roche Ltd, Basel, Switzerland}

\maketitle

\begin{abstract} {\color{black} When planning an oncology clinical trial, 
the
    usual approach is to assume {\color{black}proportional hazards and even}
    an exponential distribution for time-to-event endpoints. Often,
    besides the gold-standard endpoint overall survival {\color{black}(OS)},
    progression-free survival {\color{black}(PFS)} is considered as a second
    confirmatory endpoint. We use a survival multistate model to jointly model
    these two endpoints and find that neither exponential distribution nor
    proportional hazards will typically hold for both endpoints
    simultaneously. The multistate model {\color{black}provides a stochastic
      process} approach {\color{black} to model the dependency of such endpoints
      neither requiring latent failure times nor explicit dependency modelling
      such as copulae. W}e use the multistate model framework to simulate
    clinical trials with endpoints OS and PFS and show how design planning
    questions can be answered using this approach. {\color{black}In
      particular, non-proportional hazards for at least one of the endpoints
      are naturally modelled as well as their dependency to improve planning.} We
    consider an oncology trial on non-small-cell lung cancer as a motivating
    example from which we derive relevant trial design questions. We then
    illustrate how clinical trial design can be based on simulations from 
a
    multistate model. Key applications are co-primary endpoints and
    group-sequential designs. Simulations for these applications show that the
    standard simplifying approach {\color{black}may very well} lead to
    underpowered or overpowered clinical trials. Our approach is quite general
    and can be extended to more complex trial designs, further endpoints, 
and
    other therapeutic areas. {\color{black}An R package is available on
      CRAN.}}
\end{abstract}

\section{Introduction}\label{sec:intro} {\color{black} Overall survival (OS),
  the time from trial entry to death, is accepted as the gold standard primary
  endpoint for demonstrating clinical benefit of any drug in oncology clinical
  trials. It is easy to measure and interpret \cite{pazdur2008endpoints,
    kilickap2018endpoints}, but requires usually a long-term follow-up leading
  to a high number of patients and costs.  There are several alternative
  endpoints that reflect different aspects of disease or
  treatment. Progression-free survival (PFS) is a potential surrogate of OS
  and may provide clinical benefit itself. PFS, defined as time from trial
  entry to the earlier of {\color{black}(diagnosed)} progression or death,
  measures how long the patient lives before the tumor starts to regrow or
  progress. Many oncology trials evaluate the clinical benefit based on both
  OS and PFS. If at least one of the two needs to be statistically significant to claim \enquote*{trial success}, the pertinent FDA guidance \cite{FDA_2017} calls this \textit{multiple primary endpoints}, and that is the terminology we are going to use in this paper.  Commonly group-sequential trial designs are used for both endpoints in such a scenario. When doing formal hypothesis testing in 
such trials there are (at least) two potential sources of multiplicity: 
  investigations of the null hypothesis are done at different study times 
and for more than one endpoint are considered. There are various approaches for such trial
  designs and significance level control \cite{halabi2019textbook,
    klein2014handbook}. A common approach for sample size and power
  calculation is a simple Bonferroni correction, i.e.\ to split the
  significance level between the two endpoints and plan the sample sizes
  independently not exploiting the {\color{black}dependency} between OS 
and
  PFS.  Additionally, the current approach for trial design typically assumes proportional
  hazards (PH) for both endpoints. In this paper, we will discuss why this
  leads to inconsistencies.  Meller \textit{et~al.} \cite{meller2019joint}
  presented a simple multistate model (MSM) that jointly models OS and PFS
  {\color{black}and derived measures of dependence}. {\color{black}Here,
    we use the MSM approach for trial planning. Measures of dependence may be
    derived and used along the way, but this is not key. Rather, we use the
    stochastic MSM process as such to exploit dependence between endpoints to
    improve trial design.}  The MSM perspective has several advantages. Firstly,
  the MSM approach guarantees the natural order of the events without the 
need
  of latent failure times \cite{fleischer2009statistical, li2015weibull},
  making it canonical and the most parsimonious way to jointly model PFS and
  OS. In particular, it ensures PFS $\leq$ OS with probability one and allows
  for PFS=OS with positive probability. Secondly, the consideration of
  transition hazards in the MSM may help to better understand certain observed
  effects in the survival functions of the endpoints, such as a delayed
  treatment effect. Moreover, the {\color{black}dependency} between the
  endpoints can be derived from the joint distribution of the endpoints
  \cite{meller2019joint} and a simulation based on the MSM and accounting 
for
  that {\color{black}dependency} is easy to implement. {\color{black}This
    will, e.g., be illustrated by calculating both the global type I error and
    the joint power.}
  
  In this paper, we propose to plan oncology trials with the confirmatory
  endpoints OS and PFS using a simulation approach based on that MSM. Besides
  the fact that this approach is very flexible and applicable to many
  potential planning issues, we see two main advantages over the standard
  approach.  {\color{black}On the one hand, we can exploit the
    {\color{black}dependency} between OS and PFS, rather than just the
    {\color{black}dependency} between the analyses at different time points
    in group-sequential designs, to gain power while controlling type I error
    (T1E).} On the other hand, {\color{black} the MSM approach clarifies why
    assuming PH} for both PFS and OS simultaneously {\color{black}is very
    unrealistic, and the MSM provides a natural way to account for this}. 
 If
  the MSM approach is used for trial design, the planning is then based 
on
  assumptions on the transition hazards that induce properties of the survival
  functions for OS and PFS, rather than direct but unrelated assumptions about
  these survival functions directly. {\color{black}This is demonstrated 
in a
    clinical trial example, and the planning methodology is implemented in the
    R package simIDM, which is available on CRAN.}

  The remainder of this paper is organized as follows.  Section \ref{sec:surv}
  gives a short overview of survival multistate models. Section
  \ref{sec:joint} introduces the MSM for OS and PFS and discusses the PH
  assumption. Section \ref{sec:simualg} briefly presents the simulation
  algorithm for our MSM. {\color{black}A motivating data example is discussed in
    Section \ref{Sec:real}.}  In Section \ref{Sec:simstudy} simulation studies
  are performed to estimate power, required sample size and T1E in trials 
with
  PFS and OS as co-primary endpoints, including group-sequential designs. 
 The
  paper concludes with a discussion in Section \ref{Sec:simstudy}.}

\section{Survival Multistate Models} \label{sec:surv}
In contrast to the standard survival model with one single endpoint, a MSM allows for the analysis of complex survival data with any finite number 
of states and any transition between these states \cite{beyersmann2011competing}.
If no transitions out of a state are modeled, the state is called \textit{absorbing}, and \textit{transient} otherwise. \\

Let $(X_t)_{t\ge0}$ be a multistate process with finite state space
$ \mathcal{K}:= \{0,1,2,\dots,K´\}$ denoting the state where an individual $i$
is in at time $t$ and, and fulfilling the time-inhomogeneous Markov
assumption. The Markov assumption guarantees that the probability of a future
transition only depends on the state currently occupied and on time~$t$. In
terms of $(X_t)_{t\ge0}$ the transition hazards $\lambda_{lm}(t)$ from state
$l$ to state $m$ are defined via:
\begin{align*}
\lambda_{lm}(t)dt & = P(X_{t+dt}= m | X_{t-}=l) \\
  &  = P(X_{t+dt}= m| \mbox{Past},  X_{t-}=l), l \neq m,
\end{align*}
where $dt$ is the length of an infinitely small time interval.

{\color{black}Note that a \emph{homogeneous} Markov assumption, i.e.,
  constant hazards, is not required.} If the {\color{black}inhomogeneous}
Markov assumption is violated, the individual transition hazards are random
quantities through their dependence on the history. Under the assumption of
random censoring their average is the so-called partly-conditional transition
rate, whose cumulative counterpart can be consistently estimated by the
Nelson-Aalen estimator \cite{niessl2021statistical} discussed
below. {\color{black}Throughout, we will typically make an inhomogeneous
  Markov assumption, but will discuss extensions beyond the Markov case.}

The  non-parametric estimator of the  matrix of transition probabilities  
$ \mathbf{P}(s,t):= (P_{lm}(s,t)=P(X_t=m|X_s=l))_ {l,m \in \mathcal{K}}$, $s\le t\in [0,\tau]$,  is the Aalen-Johansen estimator \cite{aalen2008survival}, which generalizes the Kaplan-Meier estimator to multiple 
states: 
\begin{equation}
  \label{eq:AJ}
  \widehat{\mathbf{P}}(s,t)=\prod_{s < u \leq t}
  \left(\mathbf{I}+\Delta \hat{\mathbf{A}}(u)\right),
\end{equation}
where~$\mathbf{I}$ is the $(K+1)\times (K+1)$ identity matrix and $\prod$ 
is a finite product over all uniquely observed transition times in $(s,t]$ and $\Delta \hat{\mathbf{A}}(t)$ is the matrix of increments of the Nelson-Aalen estimator of the cumulative transition hazards with nondiagonal 
entries:
\begin{equation}
\small
\Delta \widehat{A}_{lm}(t)=\frac{\text{\# observed $l$ $\rightarrow$ $m$ transitions at $t$}}{\text{\# individuals observed in state $l$ just prior to $t$}}.
\end{equation}
Its diagonal entries are such that the sum of each row equals zero.
{\color{black}For joint modelling of PFS and OS, departures from the Markov
  property would impact the death after progression hazard (see Figure
  \ref{IDM}). These could be included in a transition hazard model, and,
  hence, our approach is not restricted by the Markov assumption, see also
  Section \ref{sec:simualg}.} {\color{black}Alternatively, partly
  conditional rate modelling could be employed \cite{niessl2021statistical},
  but this is not further pursued here.}

\section{A Joint Model for OS and PFS and the Assumption of PH}
\label{sec:joint}
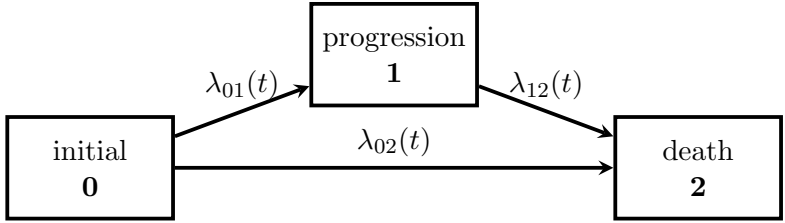
\begin{figure}

\begin{tikzpicture}[baseline=0,x=0,y=0,xscale=1,yscale=1, block/.style ={rectangle, draw=black, thick, text width=5em,align=center, minimum height=3.5em, minimum width=5.5em}]
		
		\def\x{2}
		\def\y{1}
		\def\z{0}
		\def\v{0}
	
		\node[block, line width=0.5mm] (a) at (0* \x cm, 0*\y cm) {initial \\ 
\textbf{0}};
		\node[block, line width=0.5mm] (c) at (2*\x  cm,1.5*\y cm) {progression\\ \textbf{1}};
		\node[block, line width=0.5mm] (h) at (4* \x  cm, 0*\y cm) {death \\ \textbf{2}};
		
		\draw[-stealth, line width= 0.5mm] (a) -- node [above,midway] {$\lambda_{01}(t)$} (c);	
		\draw[-stealth, line width= 0.5mm] (a) -- node [above,midway] {$\lambda_{02}(t)$} (h);		
		\draw[-stealth, line width= 0.5mm] (c) -- node [above,midway] {$\lambda_{12}(t)$} (h);	
		
		\end{tikzpicture}
		\caption{Multistate model for oncology endpoints PFS and OS}\label{IDM}
\end{figure}

We consider an illness-death model (IDM) with intermediate state
{\color{black}(diagnosed)} progression and absorbing state death to jointly
model the oncology endpoints PFS and OS (see Figure \ref{IDM}). We assume 
that
all individuals start in the initial {\color{black}s}tate 0. At the time of
the detection of disease progression, an individual makes a $0 \rightarrow 1$
transition into the progression state. Death after progression is modeled 
as a
$1 \rightarrow 2$ transition, whereas an individual who dies without prior
progression {\color{black}diagnosis} makes a direct transition from the
initial to the death state. The endpoint PFS is defined as the waiting time in
the initial state and the endpoint OS is the waiting time until the absorbing
state death is reached. PFS time is equal to OS time, if an individual dies
without prior progression.

The joint distribution of OS and PFS can be expressed by a multistate process  as \cite{meller2019joint}:
\begin{align*}
P(\mathrm{PFS} \leq u, \mathrm{OS} \leq v) & = P(X_u \in \{1, 2\}, X_v = 2)\\
&= P(X_u = 1, X_v = 2) + P(X_u = 2)\\
&= P(X_v = 2 |X_u = 1) \cdot P(X_u = 1 |X_0 = 0) \\ &+ P(X_u = 
2 |X_0 = 0), u \leq v.
\end{align*}
{\color{black}Note that dependence between PFS and OS arises naturally as a
  consequence of the MSM approach, and that the above display naturally lends
  itself to non-parametric estimation using the Aalen-Johansen estimator
  introduced earlier, not even requiring a Markov assumption under random
  censoring \cite{niessl2021statistical}.}

The OS and PFS survival functions are given by \cite{meller2019joint}:
\begin{eqnarray*}
    \spfs(t) \ := \ P(\pfs > t) &=& P_{00}(0,t), \\
     \sos(t) \ := \ P(\os > t) &=& \spfs(t) + P_{01}(0,t),
\end{eqnarray*}
{\color{black}which may {\color{black}again} be estimated by the Aalen-Johansen estimator \eqref{eq:AJ}.}

Assuming the simplest possible model for the transition hazards, namely for them to be time-constant, 
we have that \cite{meller2019joint}:

\begin{eqnarray*}
    P_{00}(0, t) &=&  \exp\Bl(-(\lambda_{01} + \lambda_{02})t\Br), \\
    P_{01}(0, t) &=& \lambda_{01} \lambda_{012}^{-1} \Bl[\exp\Bl(-(\lambda_{01} + \lambda_{02})t\Br) - \exp\Bl(-\lambda_{12}t\Br)\Br],
\end{eqnarray*}
where we abbreviated
$\lambda_{012} := \lambda_{12} - \lambda_{01} - \lambda_{02}${\color{black},
  assumed to be unequal zero,} {\color{black}and}
 $\lambda_{01}$, $\lambda_{02}$, $\lambda_{12}$ denote the respective transition-specific hazards in the IDM. 

 Using the identity $h(t)= -S(t)'/S(t)$ {\color{black}relating hazard 
and
   survival function,} the hazard function for PFS is then given by:
\begin{eqnarray}
    \hpfs(t) &=& \lambda_{01} + \lambda_{02}.
\end{eqnarray}  If we now consider two treatment arms, say $A$ and $B$,  with corresponding hazard and survival functions, then the PFS hazard ratio (HR) comparing $A$ and $B$ is as follows:
\begin{eqnarray} \label{HR_PFS}
    \hrpfs(t) &=& \frac{\lambda_{01, A} + \lambda_{02, A}}{\lambda_{01, 
B} + \lambda_{02, B}}
\end{eqnarray}  which is constant as a function of $t$, i.e.\ the PH assumption for PFS always holds for time-homogeneous transition hazards. For OS, we get \cite{meller2019joint}:
\begin{eqnarray}\label{hrOS}
    \hos(t)  &=& \frac{(\lambda_{12} - \lambda_{02})(\lambda_{01}+\lambda_{02}) - \lambda_{01}\lambda_{12}\exp(-\lambda_{012}t)}{(\lambda_{12} - 
\lambda_{02}) - \lambda_{01}\exp(-\lambda_{012}t)}.
  \end{eqnarray}
  
  It quickly becomes clear that strict assumptions for the transition hazards must be met that the quotient    $\hros(t) = \frac{h_\mathrm{OS,A}}{h_\mathrm{OS,B}}$ of two such hazards (one in each treatment group) is time-constant, i.e.\ even for the simplest model with time-constant transition hazards, the PH assumption for both PFS and OS holds only for very specific scenarios.
One assumption under which the OS hazard is time-constant, is $\lambda_{12}=\lambda_{02}$. Then, Equation \ref{hrOS}  gives $\hos(t)=\lambda_{12}$. If this is true for both groups we get a constant ratio:
\begin{eqnarray*}
    \hros(t) &=& \frac{\lambda_{12, A} }{\lambda_{12, B}}.
\end{eqnarray*}

However, the assumption  $\lambda_{12}=\lambda_{02}$  is very strong and rather unrealistic, as it implies that a progression event has no impact on the death hazard.  

Another assumption leading to PH would be $\lambda_{01}=0$ in both groups. But
that would imply that no single progression event occurs. {\color{black}
  Assuming $\lambda_{12} = \lambda_ {01} + \lambda_{02}${\color{black}, a
    truly unrealistic model, has been ruled out earlier, leading} to a
  denominator of the OS HR equal to 0.}

These considerations for the constant transition hazards make it clear that the common assumption that hazards are proportional for both endpoints 
OS and PFS is very strong and unlikely to be met in a clinical trial.

\section{Simulation of Multistate Data} \label{sec:simualg}
{\color{black}
In this section we introduce an algorithm to generate data following the IDM of Figure \ref{IDM}. Generally, arbitrary MSMs can be realized by a nested series of competing risks experiments \cite{beyersmann2011competing}. We present the algorithm for the specific case of the IDM, where we assume that all individuals start in the initial state $0$. The time dynamic algorithm helps to understand how the multistate process is constructed 
and that the transition hazards completely determine the behavior of the multistate process.
Given the transition hazards $\lambda_{01}(t)$, $\lambda_{02}(t)$ and $\lambda_{12}(t)$ the multistate trajectory for an individual can be generated as follows:
\begin{enumerate}
\item The waiting time $t_0$ in the initial {\color{black}s}tate 0 is determined by the \enquote{all-cause} hazard $\lambda_{01}(t) + \lambda_{02}(t)$, and, consequently, generated from the distribution function: $1 - 
\exp(- \int_0^t \lambda_{01}(u) + \lambda_{02}(u) \dif u)$. 
\item The state entered at $t_0$ is determined by a binomial experiment, which decides with probability $\frac{\lambda_{01}(t_0)}{\lambda_{01}(t_0) + \lambda_{02}(t_0)}$ on {\color{black}s}tate 1. If the absorbing state 2 is entered at $t_0$ the algorithm stops. Otherwise,
\item the waiting time $t_1$ in {\color{black}s}tate 1 is generated from the distribution  function: $1-\exp(- \int_{t_0}^{t_0 + s}\lambda_{12}(u) \dif u)$.
\item State 2 will be reached at time $t_0+t_1$ {\color{black}for patients
    who transit through the intermediate state}.
\end{enumerate}
Additionally, censoring times can be generated, e.g.\ by simulating random variables independent of the multistate process to obtain a random censorship model.
The algorithm assumes that the multistate process is Markov, as the transition hazards depend on the current time but the transition hazard out of 
{\color{black}s}tate 1 does not depend on the entry time $t_0$.
A non-Markov model can be simulated by modeling the transition hazard out 
of {\color{black}s}tate 1 as a function of both the entry time $t_0$ to 
the intermediate state and time since time origin in Step 3 above.
}

\section{The Oncology Trial POPLAR} \label{Sec:real}

{\color{black}
We consider a real data example, the oncology trial POPLAR \cite{rittmeyer2017atezolizumab, fehrenbacher2016atezolizumab, gandara2018blood}. The randomised open-label phase II trial POPLAR initially investigated  efficacy and safety of the anti–PD-L1 cancer immunotherapy atezolizumab 
compared to standard care (docetaxel) in patients with previously treated 
non-small-cell lung cancer.  
The primary endpoint  was OS, PFS was a secondary endpoint.

Given OS and PFS times and censoring indicators, it is easy to determine the entry and exit times of an IDM without recovery as shown in Figure \ref{IDM}.

A method to graphically check the PH assumption is to plot the cumulative
hazards for both groups against each other. If the PH assumption is true, 
then
this function should {\color{black}approximately} be a straight line through
the origin with the HR from the Cox model as approximate slope
\cite{hess1995graphical}.  Figures \ref{fig:OS_NAvsNA} and
\ref{fig:PFS_NAvsNA} show the results.  Ignoring significance testing for 
the
time being, a clear treatment effect is observed graphically in the estimated
OS survival functions, but with a relatively late separation of the treatments
(c.f.\ Figure \ref{fig:OS_NAvsNA}). Graphical evaluation of the PH assumption
for OS hazards shows that the assumption can rather be questioned (c.f.\
Figure \ref{fig:OS_NAvsNA}). The PFS survival functions cross over the course
of the trial period (c.f.\ Figure \ref{fig:PFS_NAvsNA}).
Looking at the Nelson-Aalen estimates from the corresponding IDM (see Figure   \ref{fig:hazards_NAvsNA}) it is clear, that the observed treatment effects for OS and PFS can be explained by treatment effects on the direct and post-progression death hazards, whereas no treatment effect on the progression hazard can be observed. This is in line with the understanding of how cancer immunotherapies work: The goal of immunotherapy is to boost or restore the ability of the immune system to detect and destroy cancer cells by overcoming the mechanisms by which tumors evade and suppress the immune response \cite{disis_14}. Treatment beyond progression was allowed in POPLAR, and – as opposed to chemotherapy - the patient may still benefit from the treatment even after the disease progresses. Generally, the analysis of transition hazards helps to better understand why 
the delayed treatment effect occurs.\\
It has to be noted, that we used a modified Nelson-Aalen estimator to estimate
the $1 \rightarrow 2$ cumulative transition hazard that accounts for overly
high jumps of the usual estimator at the beginning due to internal
left-truncation \cite{friedrich2017nonparametric,
  lai1991estimating}. Intermediate states in MSMs are subject to internal
left-truncation, because an individual only contributes to the risk set of a
state after its entry into that state.
Unstable estimation of the cumulative transition hazards or transition probabilities happens when an early event occurs but only a small number of 
patients are in the respective risk set, which is prevented by the methods of Lai and Ying \cite{lai1991estimating} and Friedrich \textit{et~al.}\cite{friedrich2017nonparametric}. \\
Figure \ref{fig:pw_fit} shows a parametric fit of the transition hazards using
piecewise constant hazards. We pre-specified the breaks for the constant
intervals by visual inspection and use the function \texttt{pchreg} from the
R-package \texttt{eha} \cite{eha} to fit a piecewise exponential
model. 
The parametric fit of the transition hazards is used to derive the PFS
survival functions of both groups and the result is compared with the actual
PFS survival functions in the bottom plot of Figure \ref{fig:pw_fit}, which
displays a reasonable fit after the first five time units, but would require
more refined piecewise modeling for the early shape of the PFS functions.

\begin{figure}[ht] 
\includegraphics[scale=0.55]{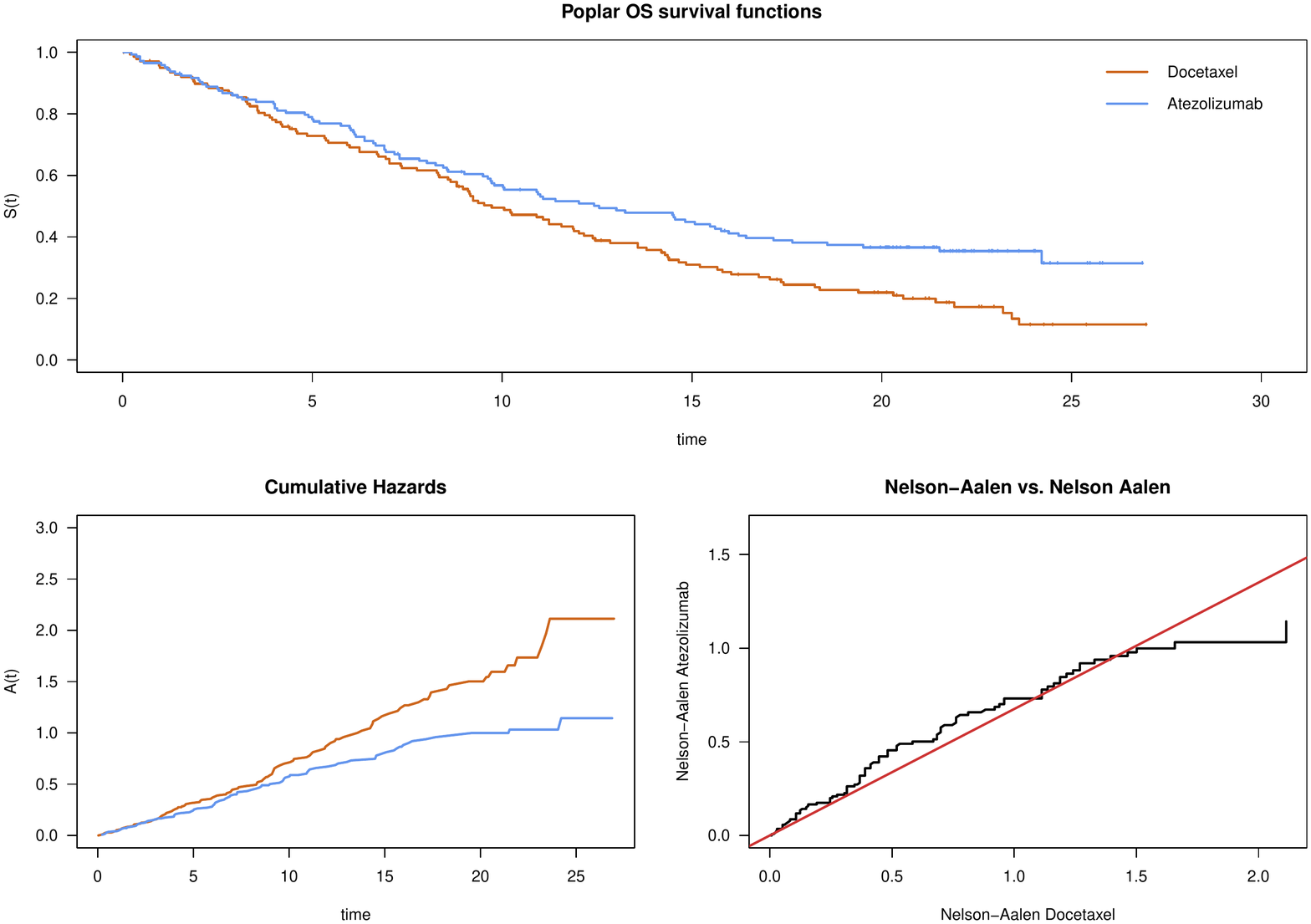}
\caption{POPLAR - Kaplan-Meier estimates of survival functions and Nelson-Aalen estimates of cumulative hazard functions, all for OS.}
\label{fig:OS_NAvsNA}
\end{figure}

\begin{figure}[ht] 
\includegraphics[scale=0.55]{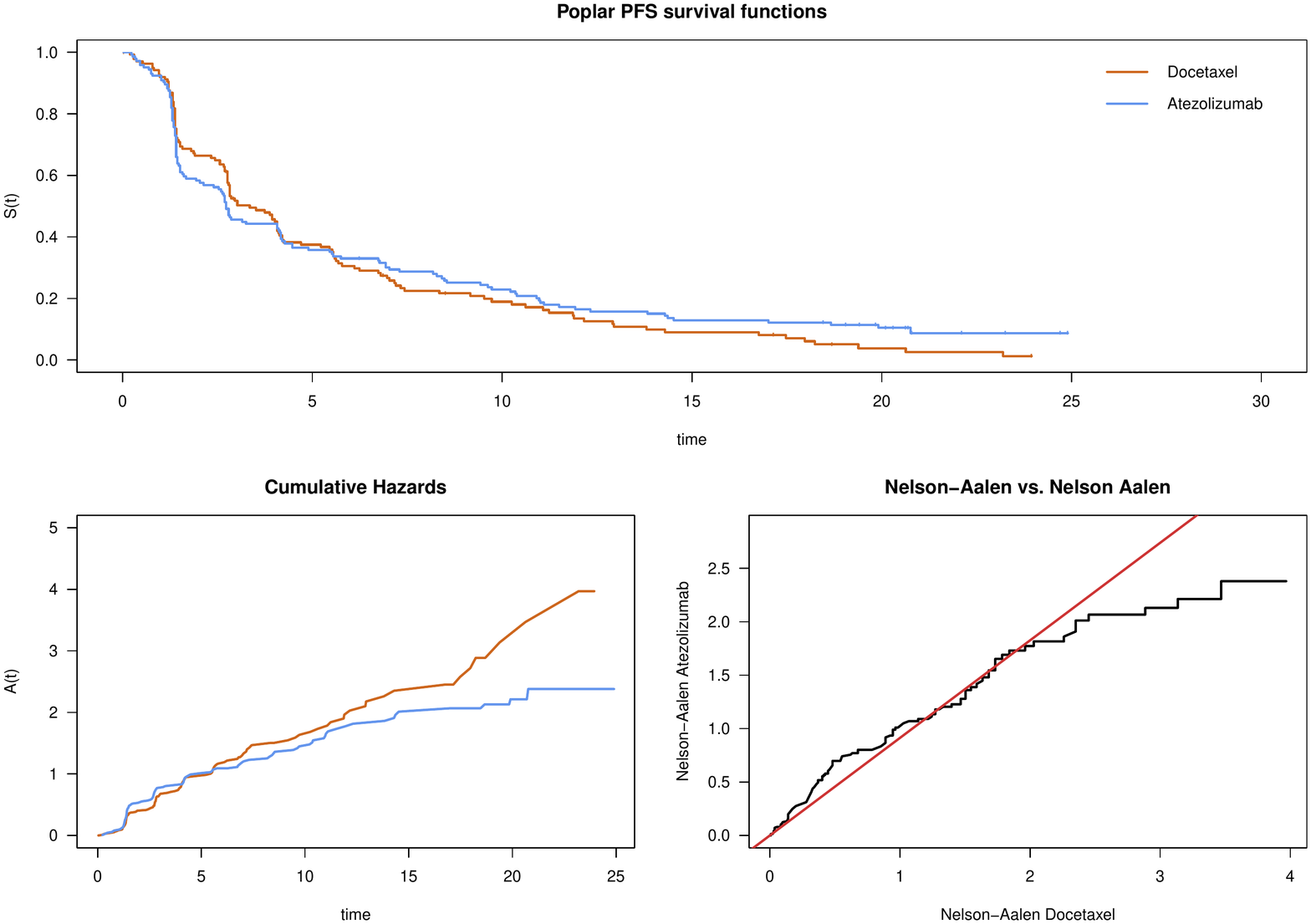}
\caption{POPLAR - Kaplan-Meier estimates of survival functions and Nelson-Aalen estimates of cumulative hazard functions, all for PFS.}
\label{fig:PFS_NAvsNA}
\end{figure}

\begin{figure}
\includegraphics[scale=0.55]{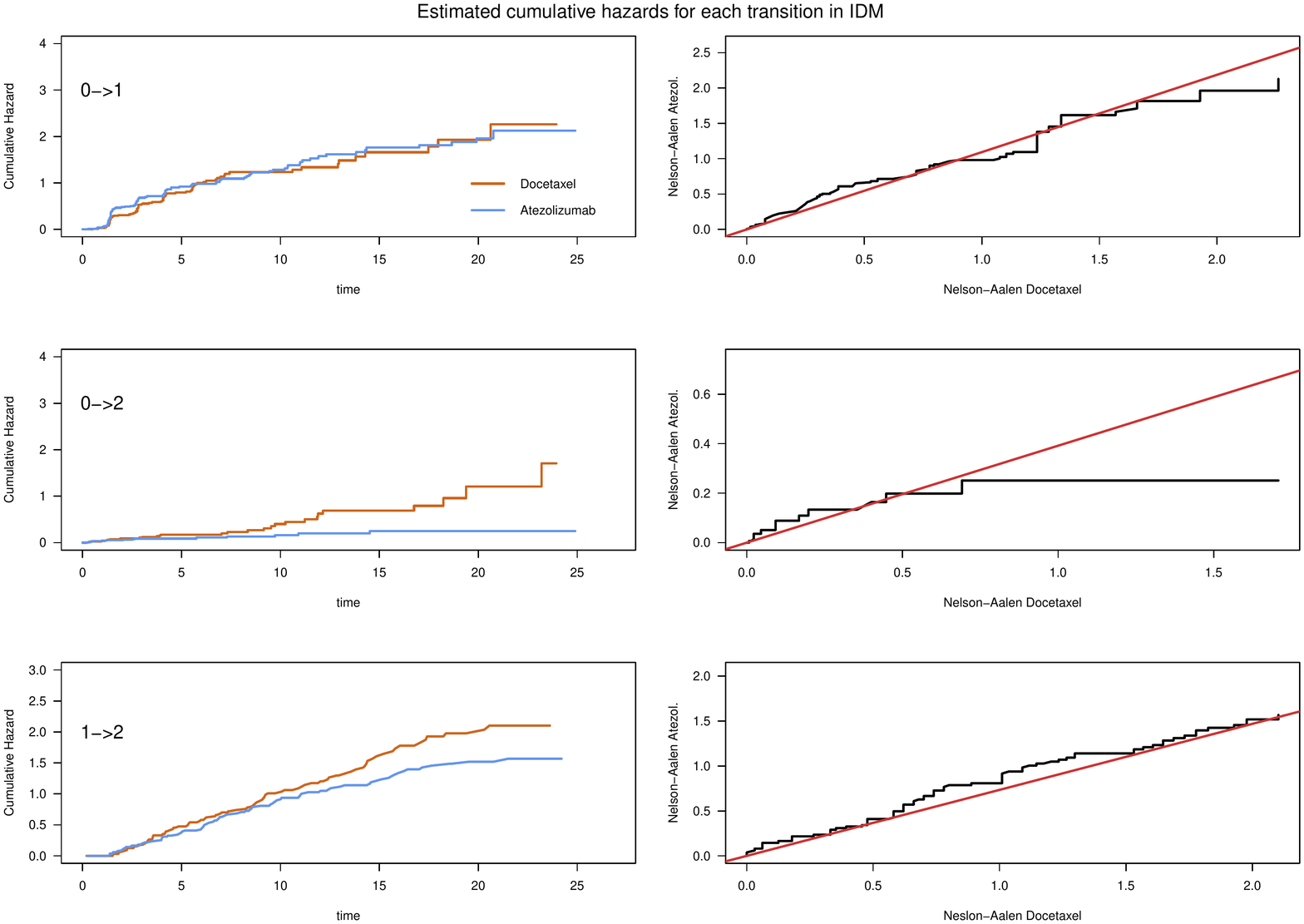}
\caption{Nelson-Aalen estimates of cumulative transition hazards (left column). Nelson-Aalen vs. Nelson-Aalen estimates (right column).}
\label{fig:hazards_NAvsNA}
\end{figure}

\begin{figure}
\includegraphics[scale=0.55]{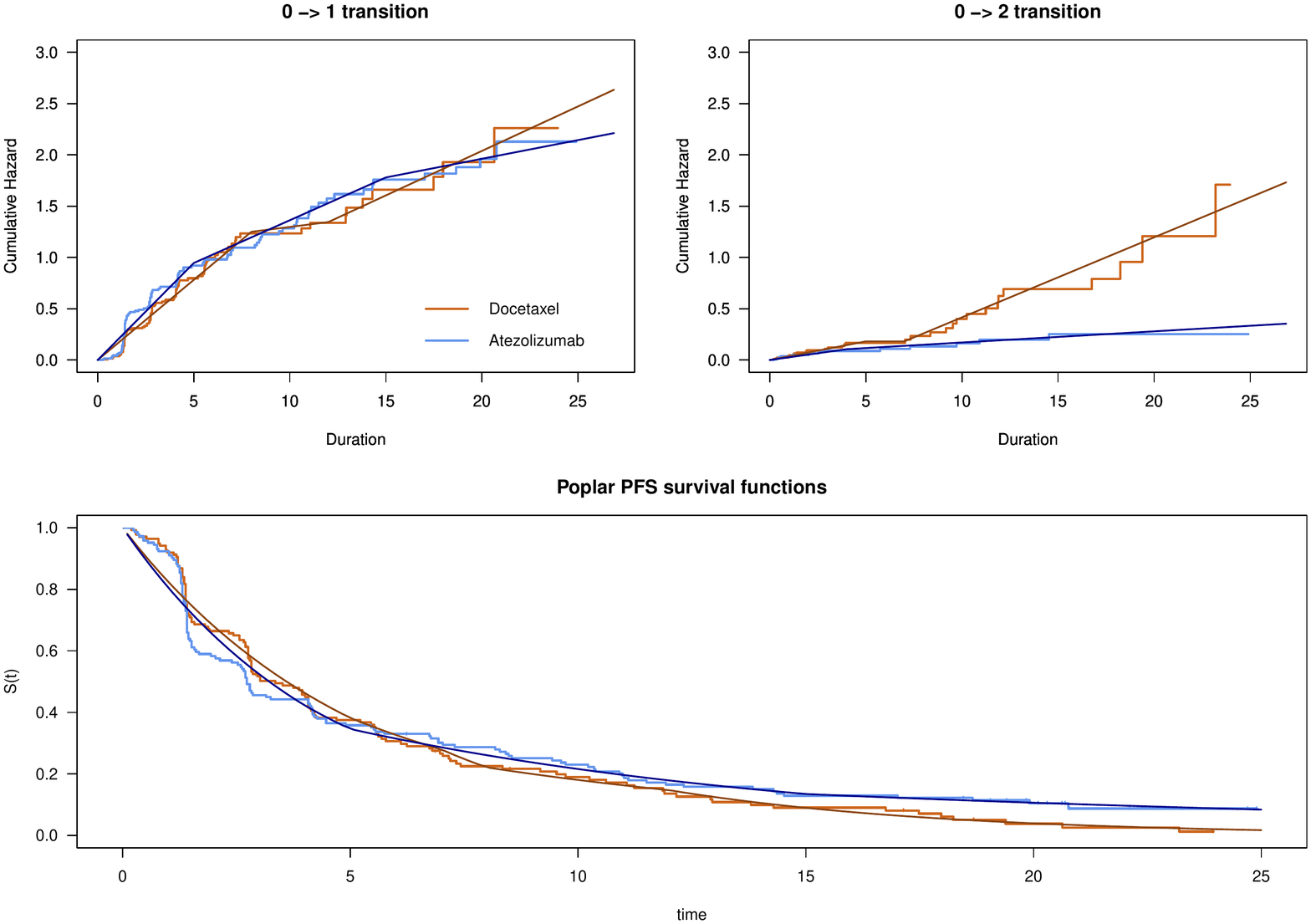}
\caption{Parametric fit of the transition hazards using piecewise constant hazards.}
\label{fig:pw_fit}
\end{figure}
 
} 

\section{Clinical Trial Design Examples} \label{Sec:simstudy} The first
clinical trial {\color{black}design} example is one with two co-primary
endpoints, OS and PFS. We compare a treatment and a control group with 1:1
randomization ratio using the assumptions in Table \ref{tab:scenarios}. For
both endpoints, a sample size calculation is performed at the planning stage
of the trial. The number of required observed events per endpoint is
determined such that a statistical power of 80\,\% for a target HR is achieved
using a log-rank test \cite{mantel1966evaluation}.  We
compare Schoenfeld's sample size calculation  \cite{schoenfeld1983sample} with simulation-based sample size calculation based on the IDM (see Table \ref{tab:scenarios}). The sample size 
by Schoenfeld is calculated separately for the two endpoints, ignoring the {\color{black}dependency} between them and assuming PH for both. The global significance level is chosen to be 5\,\%. {\color{black}A common approach to plan for co-primary endpoints is to split the global significance level} {\color{black}implying a Bonferroni correction for the two endpoints}, again not exploiting the fact that they are {\color{black}dependent}.  We specify a significance level of 4\,\% for OS and a significance level of 1\,\% for PFS. That means the probability to falsely reject the null-hypothesis of no difference between the groups for the OS endpoint is 4\,\%, and 1\,\% for the PFS endpoint. More details how to design such trial using simulation from an MSM can be found in Section \ref{sec:nointerim}. \\
In the second clinical trial design example we consider a group-sequential
design for the endpoint OS to account for the conduct of one interim
analysis. The OS interim analysis will be conducted at time of the final
analysis of the co-primary endpoint PFS. We again compare the standard
approach, namely alpha spending using the Lan-De Mets method approximating
O'Brien-Fleming boundaries \cite{demets1994interim}, to control the overall
T1E with our simulation-based approach modeling the two endpoints by an IDM.
The trial design, simulation set-up and results are discussed in Section
\ref{sec:groupseq}.  Section \ref{sec:scenarios} presents the considered
simulation scenarios to plan the two trials.

\subsection{Scenarios}\label{sec:scenarios}
\begin{figure}
\includegraphics[scale=0.55]{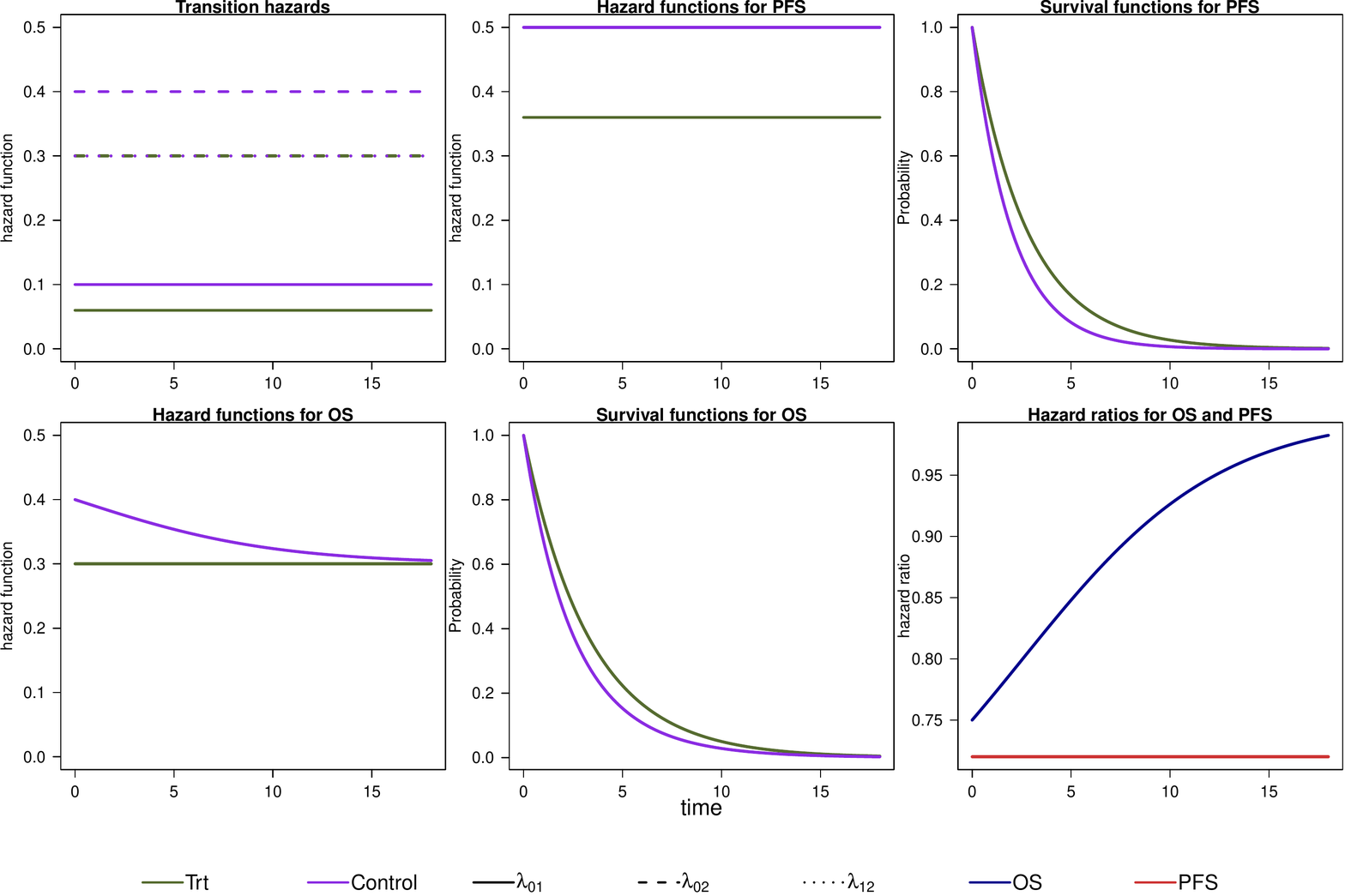}
\caption{Transition hazards, survival functions and hazard ratios for the 
Scenario 1.} 
\label{fig:sec1}
\end{figure}

\begin{figure}
\includegraphics[scale=0.55]{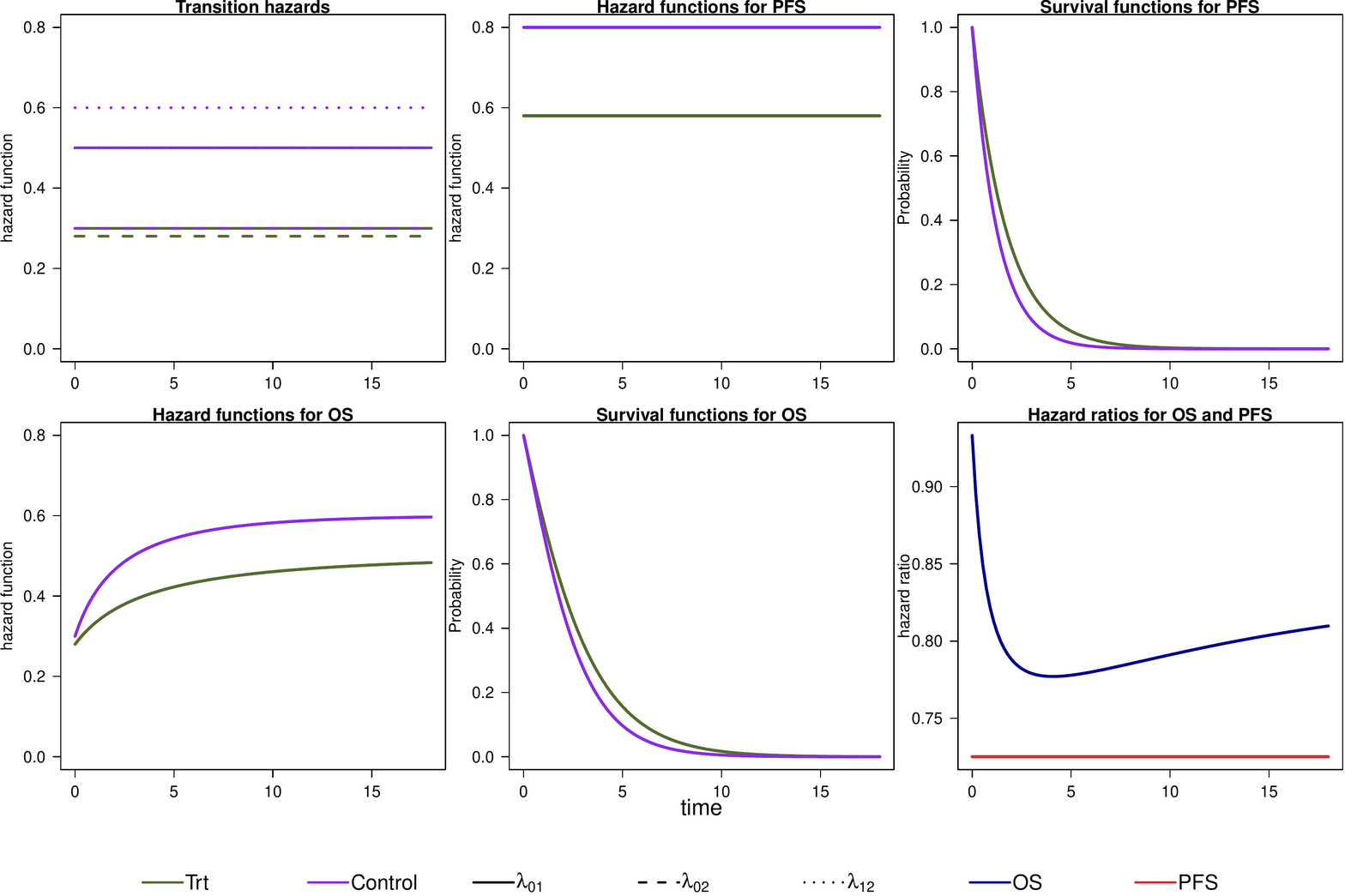}
\caption{Transition hazards, survival functions and hazard ratios for the 
Scenario 2.}
\label{fig:sec2}
\end{figure}

\begin{figure}
\includegraphics[scale=0.55]{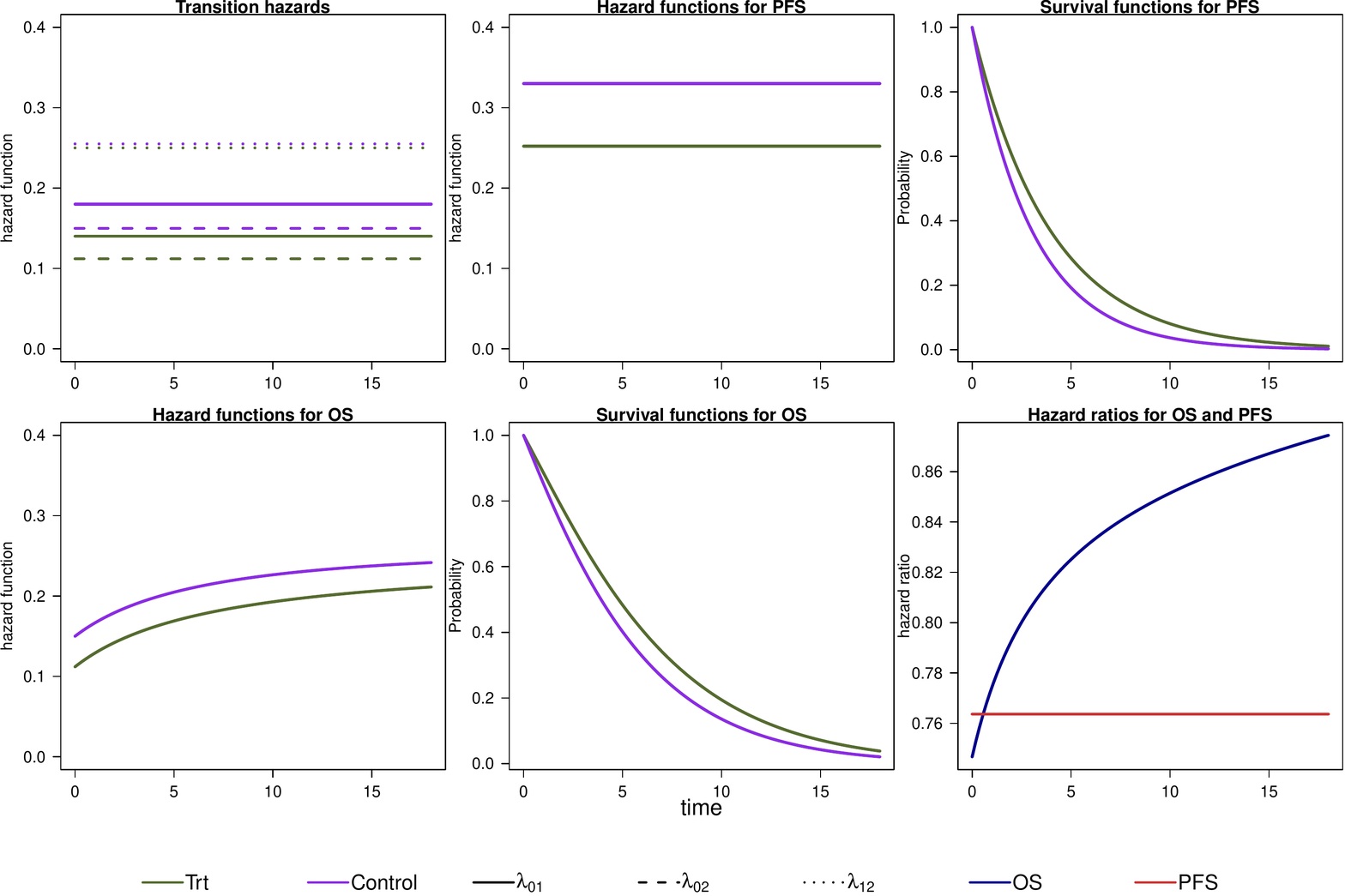}
\caption{Transition hazards, survival functions and hazard ratios for the 
Scenario 3.}
\label{fig:sec3}
\end{figure}

\begin{figure}
\includegraphics[scale=0.55]{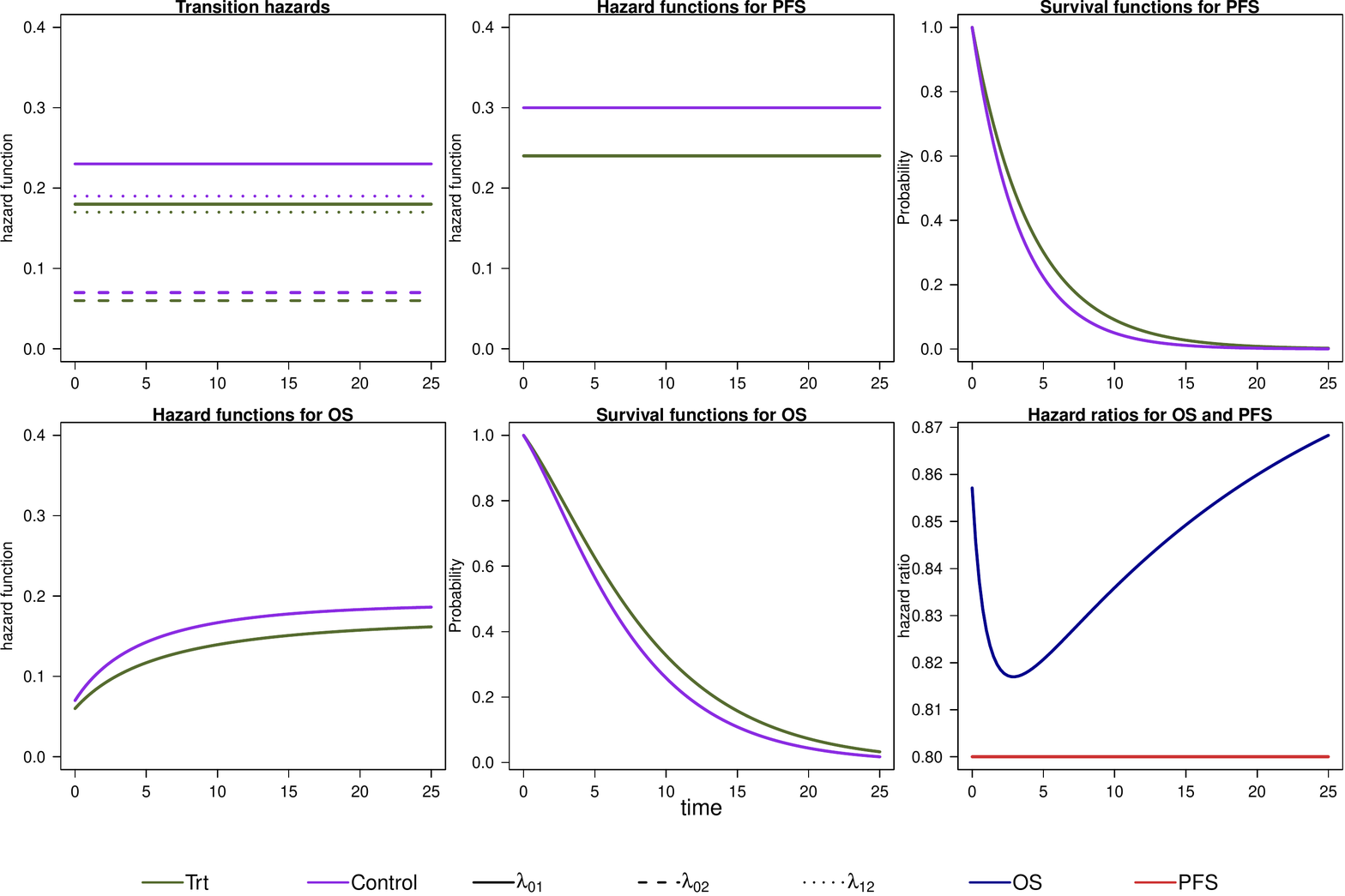}
\caption{Transition hazards, survival functions and hazard ratios for the 
Scenario 4.}
\label{fig:sec4}
\end{figure}

We model the survival functions for PFS and OS based on assumptions on the
transition specific hazards for the transitions in the IDM in Figure \ref{IDM}
in both treatment groups, i.e.\ six transition specific hazards in total. 
 We
consider four different scenarios, each with constant transition hazards for
all six transitions. The scenarios are summarized in Table
\ref{tab:scenarios}. To generate clinical trial data, for all scenarios we
simulate exponentially distributed random censoring times resulting in a
censoring probability of {\color{black} 10\,\%} within 12 time units.  In 
the
first scenario, no treatment effect in terms of $\lambda_{12}$ (i.e.\ on death
after progression) is modeled, resulting in a decreasing treatment effect 
on
OS over time (see Figure \ref{fig:sec1}).  In the second scenario we can
observe a delayed treatment effect for the OS HR because there is only a very
small treatment effect on the hazards from the initial state to death (see
Figure \ref{fig:sec2}). Scenario 3 also results in a decreasing treatment
effect over time (see Figure \ref{fig:sec3}).  In Scenario 4 the treatment
effect for OS can be explained primarily by a shorter time to disease
progression in the control group and a higher hazard for death after
progression than before progression (see Figure \ref{fig:sec4}).  Since we
consider constant transition hazards, the PFS hazards are proportional in 
all
scenarios and we know the true HR is determined by the specification of the
transition hazards in our scenarios, {\color{black}c.f.\ Equation
  \eqref{HR_PFS}}. However, we have seen in the previous sections that the PH
assumption for OS applies only to very specific scenarios, even though we 
are
considering here the simple case with constant transition hazards. In fact, in
none of our four scenarios are OS hazards proportional, see last panels in
Figures \ref{fig:sec1} - \ref{fig:sec4}. Still, {\color{black}in line with
  the common HR paradigm,} we compute a
{\color{black}time-}average{\color{black}d} HR as described by Kalbfleisch
and Prentice \cite{kalbfleisch1981estimation} from our simulated data. The PFS
HRs and the average OS HRs are displayed in Table \ref{tab:scenarios} by
scenario.

\begin{table}[ht]
\begin{center}
\begin{tabular}{ccccccc}
  \toprule
Scenario & Group & $h_{01}$ & $h_{02}$ & $h_{12}$ &PFS HR & Average OS HR 
\\
\midrule 
\multirow{2}{*}{1}& treatment group & 0.06 & 0.30 & 0.30 & \multirow{2}{*}{0.720}& \multirow{2}{*}{0.812}\\
 &  control group & 0.10 & 0.40 & 0.30 &&\\
 \midrule
\multirow{2}{*}{2} & treatment group & 0.30 & 0.28 & 0.50 & \multirow{2}{*}{0.725}& \multirow{2}{*}{0.804}\\
&  control group & 0.50 & 0.30 & 0.60 &&\\
\midrule
\multirow{2}{*}{3}& treatment group & 0.140 &  0.112& 0.250& \multirow{2}{*}{0.764}& \multirow{2}{*}{0.821}\\
&  control group & 0.180 & 0.150 & 0.255 &&\\
 \midrule
\multirow{2}{*}{4} & treatment group & 0.18 & 0.06 & 0.17 & \multirow{2}{*}{0.800}& \multirow{2}{*}{0.841}\\
 &  control group & 0.23 & 0.07 & 0.19 &&\\
 \bottomrule
\end{tabular}
\caption{Overview of scenarios. }
\label{tab:scenarios}
\end{center}
\end{table}

\subsection{Trial Design through Simulating PFS and OS as Co-Primary Endpoints - no Interim Analyses}\label{sec:nointerim}
\begin{figure}
\centering
\begin{tikzpicture}[node distance=1cm, auto]  
\tikzset{
    mynode/.style={rectangle,rounded corners,draw=black, top color=white, bottom color=red!60!black!40,very thick, inner sep=0.5em, minimum size=2em, text centered,  text width=12em},
    myarrow/.style={->, >=latex', shorten >=1pt, thick},
    mylabel/.style={text width=7em, text centered},
     mynode2/.style={rectangle,rounded corners,draw=black, top color=white, bottom color=blue!40,very thick, inner sep=0.5em, minimum size=2em, text centered},
         mynode3/.style={rectangle,rounded corners,draw=black, top color=red!60!black!40, bottom color=red!60!black!40,very thick, inner sep=0.5em, minimum size=2em, text centered, text width=6em}
      }

\node[mynode2] (manufacturer) {\begin{tabular}{c} Study design: co-primary endpoints PFS and OS \\ global significance level: $\alpha_{\mathrm{global}}=5$\,\%\end{tabular}};  
\node[text width=6em, below=1.3cm of manufacturer] (dummy) {}; 
\node[mynode, left=of dummy] (retailer1) {\begin{tabular}{c}PFS \\
$\alpha_{\mathrm{PFS}}=1$\,\% \\ 
$\mathrm{HR}_{\mathrm{PFS}}= 0.72$  \\
Power = 80\,\%\\\end{tabular}};  
\node[mynode, right=of dummy] (retailer2) {\begin{tabular}{c}OS \\
$\alpha_{\mathrm{OS}}= 4$\,\% \\
$\mathrm{HR}_{\mathrm{OS}}= 0.812$ \\
Power= 80\,\%\end{tabular}};

\node[text width=6em, below=4.5cm of manufacturer] (dummy2) {}; 
\node[mynode, left=of dummy2] (PFS) {\begin{tabular}{c}number of events: 433 \\
critical value: 2.578 \\ \end{tabular}};  
\node[mynode, right=of dummy2] (OS) {\begin{tabular}{c}number of events: 770 \\
critical value: 2.054\\ \end{tabular}};

\draw[myarrow] (retailer1) -- node [above,midway] {} (PFS);	
\draw[myarrow] (retailer2) -- node [above,midway] {} (OS);	

\node[mynode2, below=2.8cm of manufacturer] (step1) {Step 1: Schoenfeld approximation};

\node[mynode3, below=3cm of dummy2] (dummy3) {\begin{tabular}{c}  $\hat{\alpha}_{\mathrm{global}}$: 4.56
\\\end{tabular}}; 
\node[mynode, left=of dummy3] (PFS2) {\begin{tabular}{c}critical value: 2.578  \\$\hat{\alpha}_{\mathrm{PFS}}$: 1.00 
\\ \end{tabular}};  
\node[mynode, right=of dummy3] (OS2) {\begin{tabular}{c} critical value: 2.054\\$\hat{\alpha}_{\mathrm{OS}}$: 3.82
 \\\end{tabular}};

\draw[myarrow] (PFS) -- node [above,midway] {} (PFS2);	
\draw[myarrow] (OS) -- node [above,midway] {} (OS2);	

\node[mynode2, below=1cm of dummy2] (step2) {Step 2: Simulation illness-death model under $H_0$:};

%
%
%

  \node[mynode3, below=3cm of dummy3] (dummy4) {\begin{tabular}{c} joint power: \\ 71.13 \\ \end{tabular}}; 
\node[mynode, left=of dummy4] (PFS3) {\begin{tabular}{c}
\textbf{number of events: 433} \\ $\rightarrow$ empirical power: 80.94\\\end{tabular}};  
\node[mynode, right=of dummy4] (OS3) {\begin{tabular}{c} number of events: 770\\ $\rightarrow$ empirical power: 87.22\\ \textbf{number of events: 630} \\ $\rightarrow$  empirical power: 79.93 \\
 \end{tabular}};
  
 \draw[myarrow] (PFS2) -- node [above,midway] {} (PFS3);	
\draw[myarrow] (OS2) -- node [above,midway] {} (OS3);

 \node[mynode2, below=1cm of dummy3] (step4) {Step 3: Simulation under $H_1$. Get number of observed events for 80\,\% power:}; 

 \path (manufacturer) edge node  {}(retailer1)
    edge	 node {}	(retailer2);

\end{tikzpicture} 
\medskip
\caption{Simulation steps for simple study design with two co-primary endpoints. Results are shown for Scenario 1. } 
\label{diag:simple}
\end{figure}
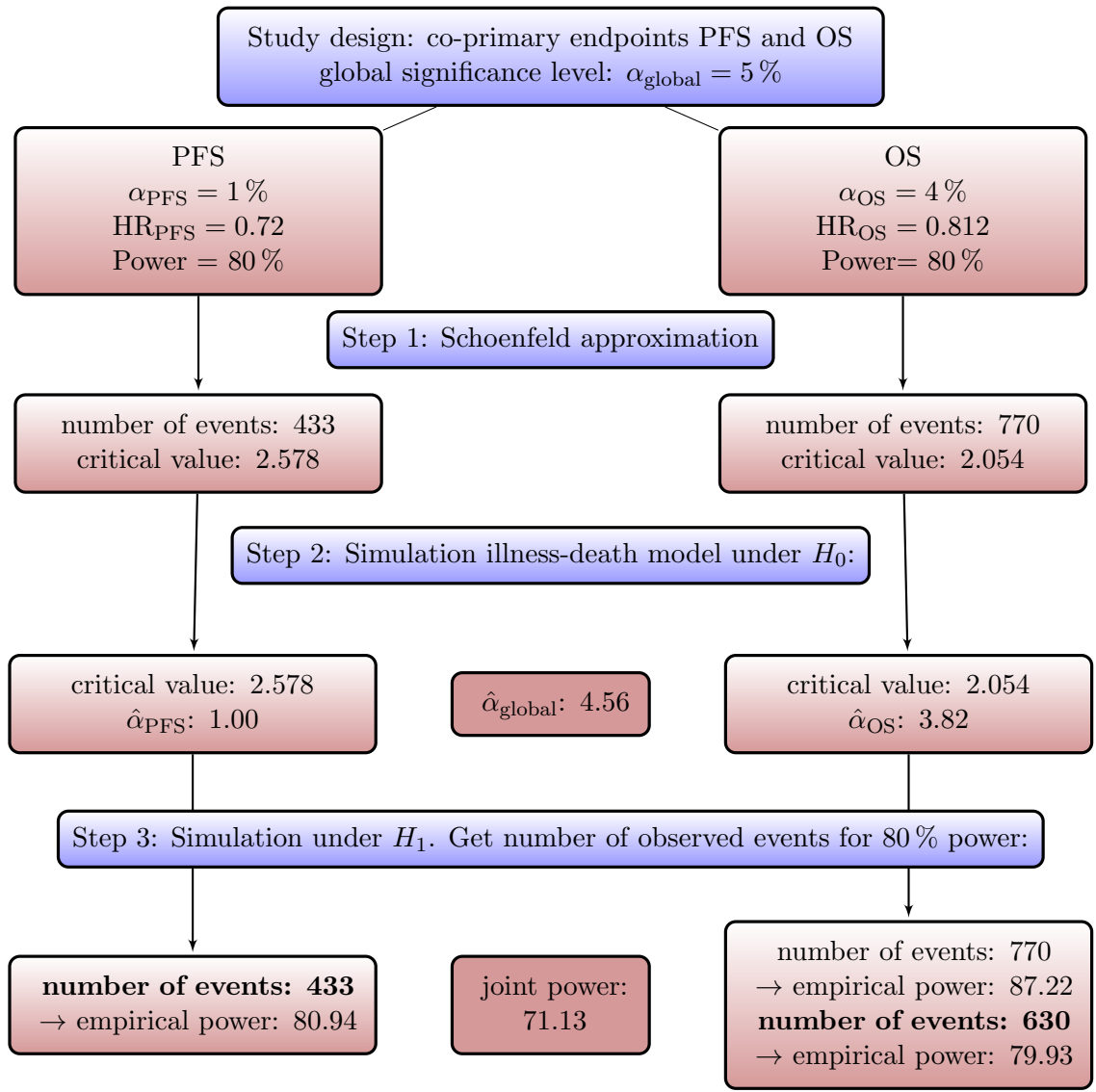

To get a starting point for calculating the required sample size via
simulation of the IDM and also to compare the simulation-based approach with
the standard approach, we calculate the sample sizes needed for OS and PFS to
get 80\,\% power to detect an improvement by the respective target HR using
the Schoenfeld's formula, respectively. For OS, we assume the effect at which we want to have 80\% power to correspond to the averaged HR for
that purpose. The global significance level of 5\,\% is split between the
co-primary endpoints. Figure \ref{diag:simple} shows the simulation steps
including the simulation results using Scenario 1 as an example. For Scenario
1, according to Schoenfeld{\color{black}'s formula} we need 433 observed PFS
events to achieve 80\,\% power at a two-sided significance level of 1\,\%
assuming a target HR of 0.72 and 770 observed OS events for 80\,\% power at a
significance level of 4\,\% and target HR of 0.812. The number of required
events for the remaining scenarios can be found in Table
\ref{tab:simuresults_simple} in the third column.  Next, we simulate the
multistate process that jointly models OS and PFS as introduced in Section
\ref{sec:joint}. The multistate process is completely determined by the
specification of the transition hazards in Table \ref{tab:scenarios}. In
Section \ref{sec:simualg} the simulation algorithm has been introduced.  The
global null hypothesis is that there is no difference between groups for both
endpoints OS and PFS, {\color{black}which means that the treatment group
  hazards are equal to those in the control group.} A false decision is made
if the null hypothesis of no difference is rejected for either OS or PFS. 
An
advantage of the joint model for OS and PFS is that the global T1E can be
estimated via simulation.  We simulate a large number (10000) of clinical
trials under the null hypothesis and determine the global T1E empirically 
by
counting the trials in which either a significant log-rank test is observed
for the OS or the PFS endpoint. Additionally, the empirical T1E probabilities
are derived for OS and PFS separately. These and the global T1E rate are shown
in columns 5 and 6 in Table \ref{tab:simuresults_simple}, see also Step 2 
in
Figure \ref{diag:simple}.  It is reasonable to suspect that we are too
conservative in splitting the significance level between the two endpoints by
simply using a Bonferroni correction to account for multiple testing of the
two endpoints, because we do not take into account the
{\color{black}dependency} of the two endpoints. {\color{black}It is possible
  to determine the critical values via simulation to fully exploit the
  significance level. Nevertheless, it is important to note that the
  significance level then relies on assumptions made during the design phase
  of the study and a design only controls T1E if the true hazard functions correspond to these assumptions.}
Next, we simulate a large number of clinical trials under $H_1$, i.e.\ we 
use
the transition hazards for the groups as specified in Table
\ref{tab:scenarios}. We use the sample sizes calculated by Schoenfeld formula
as a starting point and estimate the empirical power. That means, if the
respective number of events for an endpoint is observed, we count the number
of significant log-rank tests to obtain the empirical power for that
endpoint. Then, we increase or decrease the number of events until we obtain
the desired statistical power of approximately 80\,\%.  One advantage of the
MSM approach is, that other quantities of interest, such as joint power, can
be derived relatively easily.  The joint power is the probability that the OS
log-rank test is significant at the time of the OS analysis and,
simultaneously, the PFS log-rank test is also significant at the time of the
PFS analysis. The number of events needed for 80\,\% power can be found in
Table \ref{tab:simuresults_simple} in column 4 for all scenarios.  For
example, in Scenario 1 a sample size of 770 OS events, calculated using
Schoenfelds formula, leads to an empirical power of 88.5\,\%. For 80\,\%
power only 630 events are needed (see Step 3 in Figure
\ref{diag:simple}). {\color{black}As can be seen from
  Table~\ref{tab:simuresults_simple} sample size could be reduced with the MSM
  approach compared to the Schoenfeld approach in 3 out of 4 scenarios. This
  means, sample size calculation using Schoenfeld results in an overpowered
  trial in this scenario. In Scenario 2, more events are needed with the MSM
  approach. This means that in this scenario the Schoenfeld calculation leads
  to an underpowered trial, i.e., the risk of missing a treatment effect that
  is present is greater than planned.}

{\color{black}It should be noted that the issue is not using Schoenfeld's
  formula per se. In fact, Table~\ref{tab:simuresults_simple} illustrates
  almost perfect agreement between Schoenfeld's formula and IDM simulation
  results for PFS. The reason is that the PH assumption was valid for PFS, but
  that simple application of Schoenfeld's formula for OS does neither account
  for dependency nor for PH violation of the subsequent endpoint.}

\begin{table}[ht]
\begin{center}
\begin{tabular}{cccccc}
  \toprule
  & & \multicolumn{2}{c}{Number of observed events for 80\,\% power} &$\hat{\alpha}_{\mathrm{PFS}}$ &  \\
Scenario & Endpoint & Schoenfeld &IDM &  $\hat{\alpha}_{\mathrm{OS}}$& $\hat{\alpha}_{\mathrm{global}}$\\
\midrule 
\multirow{2}{*}{1} & PFS& 433&433&1.0& \multirow{2}{*}{4.56}\\
&OS& 770 &630&3.82&\\
\midrule
\multirow{2}{*}{2} & PFS& 452&452&0.9& \multirow{2}{*}{4.68}\\
&OS&708 &747&3.95 &\\
\midrule
\multirow{2}{*}{3} & PFS& 643&644& 1.1& \multirow{2}{*}{4.62}\\
&OS&862&742&3.83&\\\midrule
\multirow{2}{*}{4} & PFS&939 &940&0.96& \multirow{2}{*}{4.67}\\
&OS& 1113&963& 3.87 &\\
 \bottomrule
\end{tabular}
\caption{Simulation results of co-primary endpoints OS and PFS. The number of observed events that is needed for 80\,\% power obtained by Schoenfeld approximation (column 3) and by simulation of the IDM (column 4) are compared. Columns 5 and 6 show the empirical significance levels obtained via simulation of the IDM.}
\label{tab:simuresults_simple}
\end{center}
\end{table}

\subsection{Trial Design through Simulation for  PFS and OS as Co-primary 
Endpoints in a Group-Sequential Design}\label{sec:groupseq}
\begin{figure}
\centering
\begin{tikzpicture}[node distance=1cm, auto]  
\tikzset{
    mynode/.style={rectangle,rounded corners,draw=black, top color=white, bottom color=red!60!black!40,very thick, inner sep=0.5em, minimum size=2em, text centered,  text width=14em},
    myarrow/.style={->, >=latex', shorten >=1pt, thick},
    mylabel/.style={text width=7em, text centered},
     mynode2/.style={rectangle,rounded corners,draw=black, top color=white, bottom color=blue!40,very thick, inner sep=0.5em, minimum size=2em, text centered},
         mynode3/.style={rectangle,rounded corners,draw=black, top color=red!60!black!40, bottom color=red!60!black!40,very thick, inner sep=0.5em, minimum size=2em, text centered, text width=6em}
      }

\node[mynode2] (manufacturer) {\begin{tabular}{c}Group-sequential design with \\ co-primary endpoints PFS and OS \\ global significance level: $\alpha_{\mathrm{global}}= 5$\,\%\end{tabular}};  
\node[text width=6em, below=1.3cm of manufacturer] (dummy) {}; 
\node[mynode, left=of dummy] (retailer1) {\begin{tabular}{c} PFS FA at time of OS IA \\
$\alpha_{\mathrm{PFS}}=1$\,\% \\
$\mathrm{HR}_{\mathrm{PFS}}= 0.72 $ \\
Power PFS = 80\,\% \end{tabular}};  
\node[mynode, right=of dummy] (retailer2) {\begin{tabular}{c} 1 IA + 1 FA \\
cumulative $\alpha_{\mathrm{OS}}=4$\,\% \\
$\mathrm{HR}_{\mathrm{OS}}= 0.812 $ \\
Power OS overall = 80\,\% \end{tabular}};

\node[text width=6em, below=5.8cm of manufacturer] (dummy2) {}; 
\node[mynode, left=of dummy2] (PFS) {\begin{tabular}{cccc}&n& $\alpha: $ &crit.\ val.\\\
FA:&433&1.0&2.576\\ 
 \end{tabular}};  
\node[mynode, right=of dummy2] (OS) {\begin{tabular}{cccc}&n& $\alpha: $ &crit.\ val.\\
IA:&310&0.05&3.498\\ 
FA:&774&3.98&2.055\\ \end{tabular}};

\draw[myarrow] (retailer1) -- node [above,midway] {} (PFS);	
\draw[myarrow] (retailer2) -- node [above,midway] {} (OS);	

\node[mynode2, below=3.2cm of manufacturer, align=left] (step1) {Step 1: Derive O'Brien-Fleming boundaries approximated using the Lan-DeMets method. \\
Assumptions: OS and PFS are independent and both fulfill the PH assumption.};

\node[mynode3, below=3.5cm of dummy2] (dummy3) {\begin{tabular}{c}  $\hat{\alpha}_{\mathrm{global}}$: 4.84 \end{tabular}}; 
\node[mynode, left=of dummy3] (PFS2) {\begin{tabular}{c}$\hat{\alpha}_{\mathrm{PFS}}$: 1.11
\end{tabular}};  
\node[mynode, right=of dummy3] (OS2) {\begin{tabular}{cc}&$\hat{\alpha}: $ \\
IA:& 0.10\\  
FA:& 4.02\\
cumulative: & 4.04 \\ \end{tabular}};

\draw[myarrow] (PFS) -- node [above,midway] {} (PFS2);	
\draw[myarrow] (OS) -- node [above,midway] {} (OS2);	

\node[mynode2, below=1.7cm of dummy2] (step2) {Step 2: Simulation of illness-death model under $H_0$:};

%
%
%
%

  \node[mynode3, below=2.6cm of dummy3] (dummy4) {\begin{tabular}{c} joint power: \\ 72.41\\ \end{tabular}}; 
\node[mynode, left=of dummy4] (PFS3) {\begin{tabular}{cc} \# events & power PFS \\ 433 & 79.96\\
\end{tabular}};  
\node[mynode, right=of dummy4] (OS3) {\begin{tabular}{ccc} \multicolumn{2}{c}{\# events}&  \\ IA & FA & power OS \\ 380 & 774& 92.42 \\ 
380 & \textbf{524} & 80.89 \\
 \end{tabular}};
 
 \draw[myarrow] (PFS2) -- node [above,midway] {} (PFS3);	
\draw[myarrow] (OS2) -- node [above,midway] {} (OS3);

 \node[mynode2, below=1cm of dummy3] (step4) {Step 3: Update number of events for IA and FA OS analyses for 80\% power under $H_1$:}; 

 \path (manufacturer) edge node  {}(retailer1)
    edge	 node {}	(retailer2);

\end{tikzpicture} 
\medskip
\caption{Simulation steps for the group-sequential design with the results of Scenario 1. IA: interim analysis. FA: final analysis.} 
\label{diag:groupseq}
\end{figure}
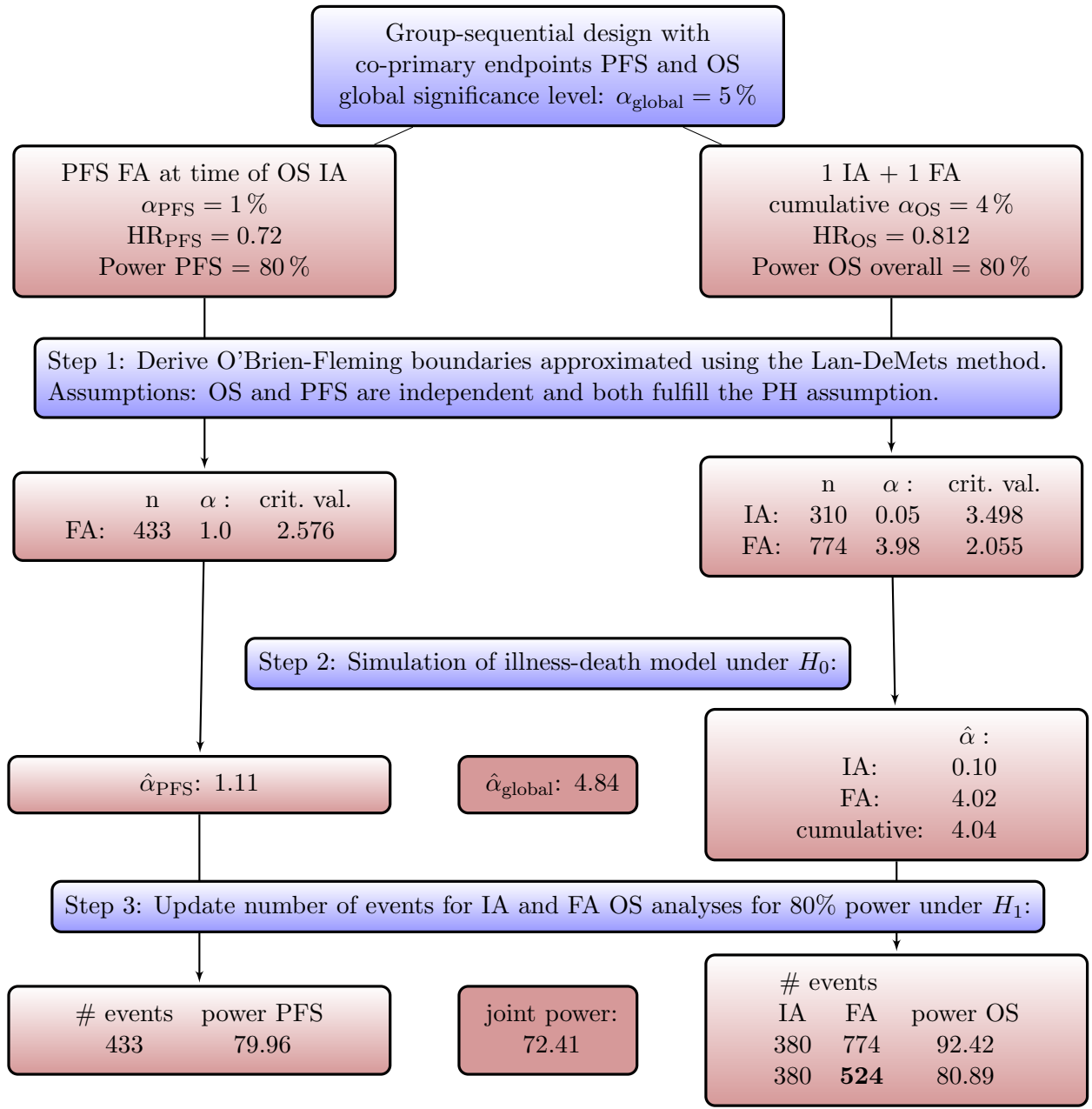

In this section, we again consider a trial design with two co-primary
endpoints PFS and OS, but we now extend the design to include an interim
analysis for OS.  This means, for the endpoint OS an interim analysis and 
a
final analysis are conducted. For PFS there is only the final analysis. The OS
interim analysis should take place at the time of the PFS final analysis. 
The
trial design and the simulation steps are shown in Figure \ref{diag:groupseq}
together with the results of Scenario 1.  To obtain a starting point for the
simulation and to have the possibility to compare the standard approach with
the simulation-based approach, we plan a group-sequential design to account
for the conduct of a OS interim analysis assuming exponentially distributed OS
and PFS survival times. An alpha spending using the Lan-De Mets method
approximating O'Brien-Fleming boundaries \cite{demets1994interim} will be
utilized to control the overall T1E probability of 4\,\% for the OS endpoint.
The R-package \texttt{R-pact} \cite{rpact} is used to plan this design (see
also Step 1 of Figure \ref{diag:groupseq}). The number of events required 
for
80\,\% power for both endpoints is listed for all scenarios in Table
\ref{tab:simuresults_groupseq} in columns 2--4. Figure \ref{diag:groupseq}
shows the critical values, number of required events and the alpha spent at
each analysis.  That standard design assumes on the one hand PH for both
endpoints and on the other hand does only account for the
{\color{black}dependency} between the multiple analyses of the OS endpoint,
but not for the {\color{black}dependency} of the PFS and OS endpoint.  Next,
we simulate a large number of clinical trials under $H_0$, i.e.\ without any
difference in the transition hazards between the group resulting in no
treatment effect for OS and PFS. We use the critical values and the timing of
the analyses as planned in Step 1 and calculate the empirical T1E
probabilities for each endpoint and each analysis as well as the cumulative
T1E of OS analyses and the global one (see also Step 2 in Figure
\ref{diag:groupseq}). In Table \ref{tab:simuresults_groupseq} the cumulative
OS T1E and the global T1E from Step 2 can be found for all scenarios.
{\color{black} As expected and as can be seen in Table
  \ref{tab:simuresults_groupseq}, the significance level is slightly
  conservative. Nevertheless, it is not decisive, as we already can observe a
  substantial power gain for some scenarios by {\color{black}jointly}
  modeling the OS and PFS survival functions.}

\begin{figure}
\includegraphics[scale=0.55]{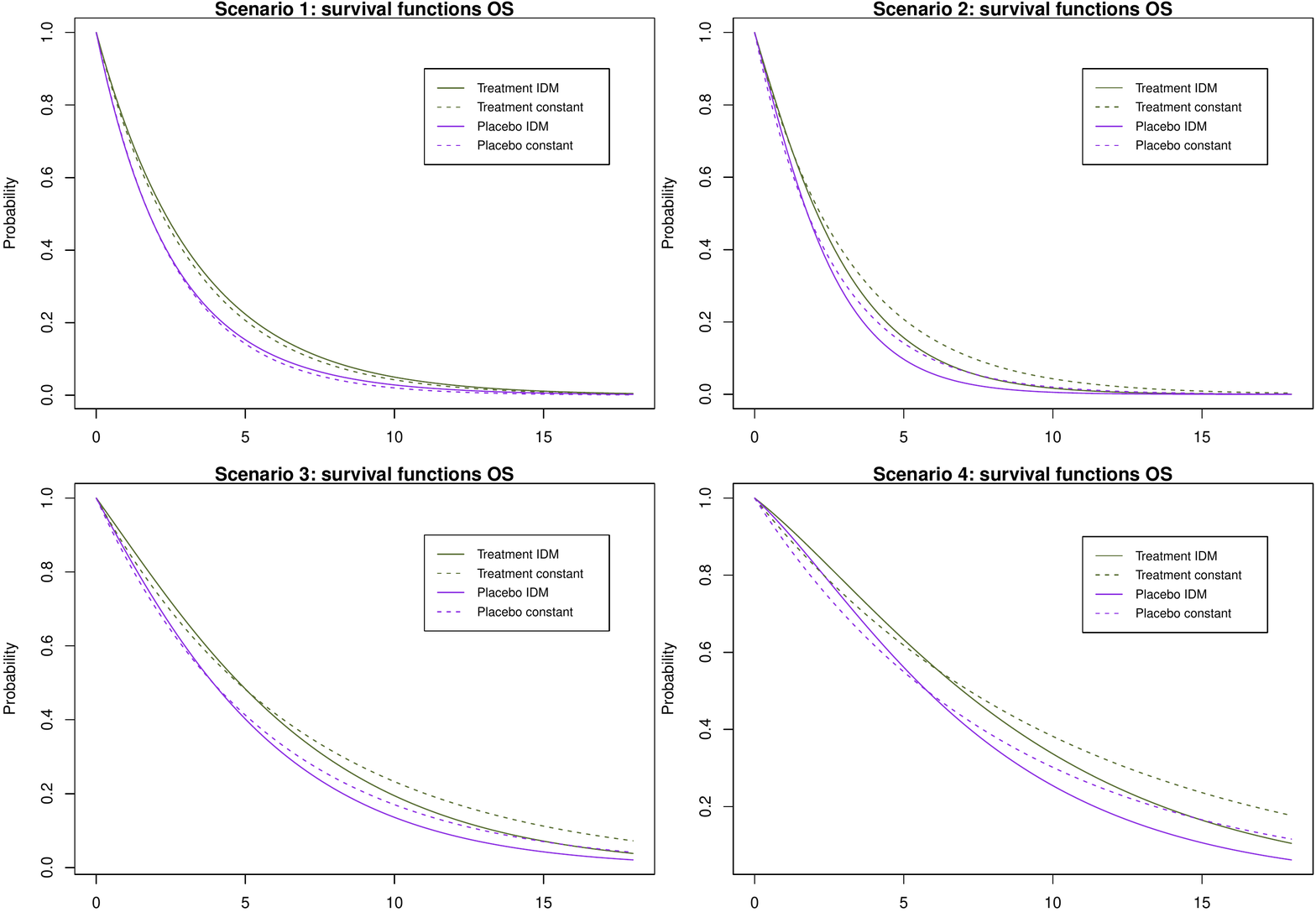}
\caption{Comparison of survival functions of OS: as specified by the IDM vs.\ fulfilling the assumption of constant OS hazards. }
\label{fig:OSSurv}
\end{figure}

The next step is to simulate the trial design under $H_1$. Our aim is to find the number of events required in the final OS analysis to achieve  80\,\% overall power for OS, and to time the interim analysis to achieve 80\,\% power for PFS as well. Similar to the results in Section \ref{sec:nointerim} we can see from Table \ref{tab:simuresults_groupseq} that in 3 out of 4 scenarios less events are needed for the final OS analysis than suggested by the O'Brien-Fleming approach. Scenario 2 shows the risk of planning an underpowered trial with the standard approach.  
Figure \ref{fig:OSSurv} shows the survival functions resulting from the specified transition hazards compared to the OS survival functions resulting from exponentially distributed survival times, that are used as input for trial design planning with \texttt{R-pact}. For the exponentially distributed OS survival times constant hazards are chosen such that the resulting median OS survival time is equal to the median OS survival time resulting from the MSM specification.


\begin{table}[ht]
\begin{center}
\begin{tabular}{c|ccc|ccc|cc}
  \toprule
   & \multicolumn{6}{c}{required events for 80\,\% power} & &  \\
Scenario & \multicolumn{3}{c} {O'Brien-Fleming} &\multicolumn{3}{c}{IDM} &&\\
& PFS & OS IA & OS FA & PFS & OS IA & OS FA &  overall $\hat{\alpha}_{\mathrm{OS}}$& $\hat{\alpha}_{\mathrm{global}}$ \\
\midrule 
1 & 433 & 310 & 774 & 433 & 380 & 524 &4.04& 4.84 \\
\midrule
2 & 452 & 212 &705 & 452 &  318 & 826& 4.08&4.80 \\
\midrule
3 & 644 & 346 & 863 & 644 & 432 &613&4.01&4.74\\
\midrule
4 & 938 & 279& 1131& 938& 502 & 919 &4.04&4.81\\

 \bottomrule
\end{tabular}
\caption{Simulation results of co-primary endpoints OS and PFS. The number of observed events that is needed for 80\,\% power obtained by O'Brien-Fleming approach (columns 2 -- 4) and by simulation of the IDM (columns 5 
-- 7) are compared. Columns 8 and 9 show the empirical significance levels obtained via simulation of the IDM.}
\label{tab:simuresults_groupseq}
\end{center}
\end{table}

\section{Discussion}\label{sec:discuss} {\color{black} In this paper we
  provided the framework to apply a MSM approach for trial planning in
  oncology. MSMs are a natural and flexible way to model complex courses of
  disease and to gain a better understanding of the data. While we focused
  here on PFS and OS, which are standard endpoints in oncology, the general
  concept and the imposed potential power gains are equally applicable in
  other therapeutic areas where the association between clinical endpoints can
  be modeled through a MSM. MSMs are also very well suited to represent trials
  with recurrent events \cite[see e.g.][]{andersen2019modeling}. For example,
  multiple sclerosis endpoints could be a further application \cite{buehler}.
  Although standard oncology endpoints as OS and PFS are often analyzed quite
  independently of each other, there are several multistate modeling
  approaches proposed for analyzing oncology endpoints or decision making 
in
  oncology \cite[e.g.][]{putter2006estimation, danzer2022confirmatory,
    beyer2020multistate}.  In this paper, we focused on how to simulate
  clinical trial data with OS and PFS endpoints and use these simulations 
for
  T1E and power
  estimations. 
Use of simulations for trial design  gains acceptance by health authorities \cite{FDA}.
  Our simulations have the great advantage that we can account for the
  {\color{black}dependency} between the two endpoints without the need of a
  closed form solution of the joint distribution of OS and PFS.\\ We
  considered a simple standard trial designs with PFS and OS as co-primary
  endpoints and a group-sequential design with one interim analysis for OS. We
  are convinced that there are a large number of possible planning questions
  on many different designs that could be answered with our approach, because
  many oncology trials are planned taking into account both PFS and OS.
  Whereas, of course, some further research is needed to provide detailed
  guidance for specific trial designs, this paper provides the tools and the
  framework to address planning questions for trial designs via simulation and
  based on a MSM. At the same time, this manuscript shows that the multistate
  approach avoids inconsistent assumptions by modeling OS and PFS together. We
  have made clear that it is a very unrealistic assumption that OS and PFS
  both satisfy the PH assumption,  {\color{black}while}  trials are {\color{black}arguably} planned making this simplifying assumption. Moreover, the MSM approach exploits the {\color{black}dependency} between the 
two endpoints. This is routinely done in group-sequential designs for the 
{\color{black}dependency} over time and also in enrichment designs  for 
the {\color{black}dependency} between nested subgroups \cite{jenkins2011adaptive}, but usually not for different oncology endpoints.\\
  {\color{black}Our IDM approach offers a way to mechanistically generate 
PFS and OS data implying a non-PH structure for OS. Instead of modeling the non-PH structure of OS on the level of survival functions, as e.g.\ done in Ristl \textit{et~al.}\cite{ristl2021delayed}, the IDM offers an approach that may be more natural and transparent with respect to how the survival functions develop. Similarly, instead of assuming non-PH on level of OS survival functions that is generated by heterogeneous treatment effects in subpopulations, we can, {\color{black}even} in a model with constant hazards for all transitions in all subgroups, simply average these hazards. As a consequence, non-PH is not observed because of heterogeneous subgroups, but indeed again because we properly model the process of generation of PFS and OS through an IDM.} \\
  Interesting metrics depending on both endpoints as the joint power and
  global T1E can also easily be derived using simulation and the multistate
  modeling framework.\\Jung \textit{et~al.} \cite{jung2018reconsideration}
  also emphasize that the PH assumption for OS is rather unrealistic. However,
  they do not consider MSMs and their approach to the problem involves strong
  assumptions about the time after progression.
  \\
  {\color{black}In our simulation studies, we have seen that there is a large potential to save sample size, but it is also possible that more events are required with the MSM depending on how the OS survival functions relate and how  much the PH assumption is violated. Nevertheless, planning based on the MSM reflects a more realistic power estimation, which is of course in the interest of the trial planners. It should be emphasized that our recommendation is to always plan trials with the IDM approach and 
not to pick the standard approach in situations like Scenario 2 in Table \ref{tab:simuresults_groupseq}, because the simplifying assumption of PH for OS is simply misleading (see Figure \ref{fig:OSSurv}).  That means planning a trial by simulating an IDM may potentially offer a very relevant 
efficiency gain, e.g.\ by reducing the number of OS events at which the OS final analysis is performed quite dramatically (Scenarios 1, 3, 4 in Table \ref{tab:simuresults_groupseq}) or guard against an underpowered OS analysis (in Scenario 2). How can these patterns be explained? Comparing the time-dependent OS HRs in Figures \ref{fig:sec1} - \ref{fig:sec4} (right bottom panel) we see that for Scenarios 1, 3, 4 the OS effect is increasing over time while for Scenario 2 it is decreasing. Further research is 
needed to understand these patterns, as we need to consider two factors that influence the power. The IDM accounts for the {\color{black}dependency} between the two endpoints, which is expected to result in a power gain, at the same time the PH assumption is not met for OS, which might result in a power loss when applying the log-rank test, depending on the shape of the OS survival functions induced through the IDM.}\\
  To detect a treatment difference between the groups we used the standard log-rank test, which can lead to a power loss if  non-PH are present \cite{lagakos1984properties}. The log-rank test can be modified such that individual weights are assigned to the events \cite{harrington1982class, kalbfleisch2011statistical} or by combining several pre-specified weight functions and accounting for the {\color{black}dependency} between the resulting test statistics \cite[max-combo tests, see e.g.][]{lin_20}. Another common analysis method in the presence of non-PH is the calculation of restricted mean survival (RMST) \cite{royston2013restricted}.  It is of 
course possible to use a different statistical test in our MSM approach. Whether this leads to a gain of power is a target of future research. \\
  It is important to note that the trial planning with the MSM approach is based on the assumptions of the transition hazards instead of OS and PFS survival functions. The transition hazards are more flexible in that they can be  chosen independently for each transition  (while OS and PFS hazards depend on each other) and certain effects can be better understood, 
but the specification requires a shift in thinking. Identifying planning assumptions might also require re-estimating  transition-specific hazards 
from earlier trials to inform planning hazards such that they match PFS and OS survival functions.\\
  In this paper we considered a MSM that fulfills the Markov
  assumption. However, the IDM allows to explicitly model and simulate a
  departure of the Markov assumption. Therefore, our approach is also
  applicable to non-Markov situations, e.g.\ if the time of progression
  influences subsequent survival.  Meller
  \textit{et~al.}\cite{meller2019joint} also discussed the joint distribution
  of PFS and OS in the non-Markov case. Theoretically, the Markov assumption
  leads to a greater efficiency of the Aalen-Johansen estimator compared to
  the Kaplan-Meier estimator \cite{andersen2012statistical}. 
  To what extent this is relevant for the power calculations in our context could be the subject of future research. \\
  We also implemented an R-package \texttt{simIDM} \cite{simIDM} that can 
be
  used to simulate a large number of clinical trials with endpoints OS and
  PFS. The package provides several features, besides constant transition
  hazards, IDMs based on Weibull hazards or piecewise constant hazards can be
  simulated as well. Moreover, both random censoring and event-driven
  censoring \cite{ruhl2022general} as well as staggered study entry can be
  specified. The number of trial arms and the randomization ratio can also be
  varied.  }

\section{Acknowledgments}
The authors would like to thank Daniel Sabanes Bove for his help in developing
the R-package \texttt{simIDM} \cite{simIDM} used for the simulations in this
paper. {\color{black}Jan Beyersmann acknowledges support from the German Research
  Foundation (DFG grant BE 4500/6-1).}
\section{Data Availability Statement}
{\color{black}The data presented in this study are openly available at  \url{https://www.nature.com/articles/s41591-018-0134-3#Sec19} (Supplementary Table 8).

}

\section{Conflicts of Interest}
The authors declared no potential conflicts of interest with respect to the research, authorship and/or publication of this article.
\bibliographystyle{wiley}
\bibliography{simengine.bib}

\begin{thebibliography}{10}
\providecommand{\url}[1]{\texttt{#1}}
\providecommand{\urlprefix}{URL }
\expandafter\ifx\csname urlstyle\endcsname\relax
  \providecommand{\doi}[1]{doi:\discretionary{}{}{}#1}\else
  \providecommand{\doi}{doi:\discretionary{}{}{}\begingroup
  \urlstyle{rm}\Url}\fi

\bibitem{pazdur2008endpoints}
Pazdur R. Endpoints for assessing drug activity in clinical trials. \emph{The
  Oncologist}  2008; \textbf{13}(S2):19--21.

\bibitem{kilickap2018endpoints}
Kilickap S, Demirci U, Karadurmus N, Dogan M, Akinci B, Sendur MAN. Endpoints
  in oncology clinical trials. \emph{J buon}  2018; \textbf{23}(7):1--6.

\bibitem{FDA_2017}
{US Food and Drug Administration}. \emph{Guidance for Industry: Multiple
  Endpoints in Clinical Trials} 2017.
  \urlprefix\url{https://www.fda.gov/media/102657/download}.

\bibitem{halabi2019textbook}
Halabi S, Michiels S. \emph{Textbook of clinical trials in oncology: a
  statistical perspective}. CRC Press, 2019.

\bibitem{klein2014handbook}
Ohneberg K, Schumacher M. Handbook of survival analysis. \emph{ed. by J.P.
  Klein et al. Chapman \& Hall/CRC. Chap. Sample Size Calculation for Clinical
  Trials}  2013; :571--594.

\bibitem{meller2019joint}
Meller M, Beyersmann J, Rufibach K. Joint modeling of progression-free and
  overall survival and computation of correlation measures. \emph{Statistics in
  Medicine}  2019; \textbf{38}(22):4270--4289.

\bibitem{fleischer2009statistical}
Fleischer F, Gaschler-Markefski B, Bluhmki E. A statistical model for the
  dependence between progression-free survival and overall survival.
  \emph{Statistics in Medicine}  2009; \textbf{28}(21):2669--2686.

\bibitem{li2015weibull}
Li Y, Zhang Q. A {W}eibull multi-state model for the dependence of
  progression-free survival and overall survival. \emph{Statistics in Medicine}
   2015; \textbf{34}(17):2497--2513.

\bibitem{beyersmann2011competing}
Beyersmann J, Allignol A, Schumacher M. \emph{Competing risks and multistate
  models with R}. Springer Science \& Business Media, 2012.

\bibitem{niessl2021statistical}
Nie{\ss}l A, Allignol A, Beyersmann J, Mueller C. Statistical inference for
  state occupation and transition probabilities in non-{M}arkov multi-state
  models subject to both random left-truncation and right-censoring.
  \emph{Econometrics and Statistics}  2021; \doi{10.1016/j.ecosta.2021.09.008}.

\bibitem{aalen2008survival}
Aalen O, Borgan O, Gjessing H. \emph{Survival and event history analysis: a
  process point of view}. Springer Science \& Business Media, 2008.

\bibitem{rittmeyer2017atezolizumab}
Rittmeyer A, Barlesi F, Waterkamp D, Park K, Ciardiello F, Von~Pawel J, Gadgeel
  SM, Hida T, Kowalski DM, Dols MC, \emph{et~al.}. Atezolizumab versus
  docetaxel in patients with previously treated non-small-cell lung cancer
  ({OAK}): a phase 3, open-label, multicentre randomised controlled trial.
  \emph{The Lancet}  2017; \textbf{389}(10066):255--265.

\bibitem{fehrenbacher2016atezolizumab}
Fehrenbacher L, Spira A, Ballinger M, Kowanetz M, Vansteenkiste J, Mazieres J,
  Park K, Smith D, Artal-Cortes A, Lewanski C, \emph{et~al.}. Atezolizumab
  versus docetaxel for patients with previously treated non-small-cell lung
  cancer ({POPLAR}): a multicentre, open-label, phase 2 randomised controlled
  trial. \emph{The Lancet}  2016; \textbf{387}(10030):1837--1846.

\bibitem{gandara2018blood}
Gandara DR, Paul SM, Kowanetz M, Schleifman E, Zou W, Li Y, Rittmeyer A,
  Fehrenbacher L, Otto G, Malboeuf C, \emph{et~al.}. Blood-based tumor
  mutational burden as a predictor of clinical benefit in non-small-cell lung
  cancer patients treated with atezolizumab. \emph{Nature Medicine}  2018;
  \textbf{24}(9):1441--1448.

\bibitem{hess1995graphical}
Hess KR. Graphical methods for assessing violations of the proportional hazards
  assumption in cox regression. \emph{Statistics in Medicine}  1995;
  \textbf{14}(15):1707--1723.

\bibitem{disis_14}
Disis ML. Mechanism of action of immunotherapy. \emph{Seminars in Oncology}
  2014; \textbf{41}:S3 -- S13,
  \doi{https://doi.org/10.1053/j.seminoncol.2014.09.004}. Immuno-oncology Comes
  of Age.

\bibitem{friedrich2017nonparametric}
Friedrich S, Beyersmann J, Winterfeld U, Schumacher M, Allignol A.
  Nonparametric estimation of pregnancy outcome probabilities. \emph{The Annals
  of Applied Statistics}  2017; :840--867.

\bibitem{lai1991estimating}
Lai TL, Ying Z. Estimating a distribution function with truncated and censored
  data. \emph{The Annals of Statistics}  1991; :417--442.

\bibitem{eha}
Broström G. \emph{eha: Event History Analysis} 2020.
  \urlprefix\url{https://cran.r-project.org/package=eha}, r package version
  2.8.3.

\bibitem{mantel1966evaluation}
Mantel N. Evaluation of survival data and two new rank order statistics arising
  in its consideration. \emph{Cancer Chemotherapy Reports}  1966;
  \textbf{50}:163--170.

\bibitem{schoenfeld1983sample}
Schoenfeld DA. Sample-size formula for the proportional-hazards regression
  model. \emph{Biometrics}  1983; :499--503.

\bibitem{demets1994interim}
Demets DL, Lan KG. Interim analysis: the alpha spending function approach.
  \emph{Statistics in Medicine}  1994; \textbf{13}(13-14):1341--1352.

\bibitem{kalbfleisch1981estimation}
Kalbfleisch JD, Prentice RL. Estimation of the average hazard ratio.
  \emph{Biometrika}  1981; \textbf{68}(1):105--112.

\bibitem{rpact}
Wassmer G, Pahlke F. \emph{rpact: Confirmatory Adaptive Clinical Trial Design
  and Analysis} 2021. \urlprefix\url{https://CRAN.R-project.org/package=rpact},
  r package version 3.1.0.

\bibitem{andersen2019modeling}
Andersen PK, Angst J, Ravn H. Modeling marginal features in studies of
  recurrent events in the presence of a terminal event. \emph{Lifetime Data
  Analysis}  2019; \textbf{25}(4):681--695.

\bibitem{buehler}
Bühler A, Wolbers M, Model F, Wang Q, Belachew S, Manfrini M, Lorscheider J,
  Kappos L, Beyersmann J. Recurrent disability progression endpoints in
  multiple sclerosis clinical trials. \emph{Multiple Sclerosis Journal}  2022;
  \doi{10.1177/13524585221125382}.

\bibitem{putter2006estimation}
Putter H, van~der Hage J, de~Bock GH, Elgalta R, van~de Velde CJ. Estimation
  and prediction in a multi-state model for breast cancer. \emph{Biometrical
  Journal: Journal of Mathematical Methods in Biosciences}  2006;
  \textbf{48}(3):366--380.

\bibitem{danzer2022confirmatory}
Danzer MF, Terzer T, Berthold F, Faldum A, Schmidt R. Confirmatory adaptive
  group sequential designs for single-arm phase {II} studies with multiple
  time-to-event endpoints. \emph{Biometrical Journal}  2022;
  \textbf{64}(2):312--342.

\bibitem{beyer2020multistate}
Beyer U, Dejardin D, Meller M, Rufibach K, Burger HU. A multistate model for
  early decision-making in oncology. \emph{Biometrical Journal}  2020;
  \textbf{62}(3):550--567.

\bibitem{FDA}
{Food and Drug Administration}. Interacting with the {FDA} on complex
  innovative trial designs for drugs and biological products. \emph{Guidance
  for Industry}  2020; .

\bibitem{jenkins2011adaptive}
Jenkins M, Stone A, Jennison C. An adaptive seamless phase {II/III} design for
  oncology trials with subpopulation selection using correlated survival
  endpoints. \emph{Pharmaceutical Statistics}  2011; \textbf{10}(4):347--356.

\bibitem{ristl2021delayed}
Ristl R, Ballarini NM, G{\"o}tte H, Sch{\"u}ler A, Posch M, K{\"o}nig F.
  Delayed treatment effects, treatment switching and heterogeneous patient
  populations: How to design and analyze rcts in oncology. \emph{Pharmaceutical
  Statistics}  2021; \textbf{20}(1):129--145.

\bibitem{jung2018reconsideration}
Jung I, Ko HJ, Rha SY, Nam CM. Reconsideration of sample size and power
  calculation for overall survival in cancer clinical trials.
  \emph{Contemporary Clinical Trials Communications}  2018; \textbf{12}:90--91.

\bibitem{lagakos1984properties}
Lagakos S, Schoenfeld D. Properties of proportional-hazards score tests under
  misspecified regression models. \emph{Biometrics}  1984; :1037--1048.

\bibitem{harrington1982class}
Harrington DP, Fleming TR. A class of rank test procedures for censored
  survival data. \emph{Biometrika}  1982; \textbf{69}(3):553--566.

\bibitem{kalbfleisch2011statistical}
Kalbfleisch JD, Prentice RL. \emph{The statistical analysis of failure time
  data}. John Wiley \& Sons, 2011.

\bibitem{lin_20}
Lin RS, Lin J, Roychoudhury S, Anderson KM, Hu T, Huang B, Leon LF, Liao JJ,
  Liu R, Luo X, \emph{et~al.}. Alternative analysis methods for time to event
  endpoints under nonproportional hazards: A comparative analysis.
  \emph{Statistics in Biopharmaceutical Research}  2020;
  \textbf{12}(2):187--198, \doi{10.1080/19466315.2019.1697738}.

\bibitem{royston2013restricted}
Royston P, Parmar MK. Restricted mean survival time: an alternative to the
  hazard ratio for the design and analysis of randomized trials with a
  time-to-event outcome. \emph{BMC Medical Research Methodology}  2013;
  \textbf{13}(1):152.

\bibitem{andersen2012statistical}
Andersen PK, Borgan O, Gill RD, Keiding N. \emph{Statistical models based on
  counting processes}. Springer Science \& Business Media, 1993.

\bibitem{simIDM}
Erdmann A, Rufibach K, Sabanes~Bove D. \emph{simIDM: Simulating Clinical Trials
  with Endpoints Progression-Free Survival and Overall Survival using an
  Illness-Death Model} 2023.
  \urlprefix\url{https://CRAN.R-project.org/package=simIDM}.

\bibitem{ruhl2022general}
R{\"u}hl J, Beyersmann J, Friedrich S. General independent censoring in
  event-driven trials with staggered entry. \emph{Biometrics}  2022; .

\end{thebibliography}

\end{document}